%% file: hp2_v3.tex
\documentclass[12pt,dvips]{article}

\usepackage{cite,epsfig}
\usepackage{color}
\input def_colors

\def\slash#1{#1\!\!\!\!/}

\def\gsim{\:\raisebox{-0.5ex}{$\stackrel{\textstyle>}{\sim}$}\:}

\def\theequation{\arabic{section}.\arabic{equation}}

\begin{document}

\begin{flushright}
MAN/HEP/2008/20\\
arXiv:0807.4167 \\
July 2008
\end{flushright}

\bigskip

\begin{center}
{\bf {\LARGE Supersymmetric Higgs Singlet Effects on\\[3mm] {\boldmath
    $B$}-Meson FCNC Observables at Large {\boldmath $\tan\beta$}}}
\end{center}

\bigskip

\begin{center}
{\large Robert N. Hodgkinson$^{a,b}$ and Apostolos Pilaftsis$^{a}$}
\end{center}

\begin{center}
{\em $\ ^a$ School of Physics and Astronomy, University of Manchester}\\
{\em Manchester M13 9PL, United Kingdom}\\[3mm]
{\em $\ ^b$ High Energy Physics Group, Dept. ECM, Univ. de Barcelona}\\
{\em Av. Diagonal 647, E-08028 Barcelona, Catalonia, Spain}
\end{center}

\bigskip\bigskip\bigskip


\centerline{\bf ABSTRACT}

\noindent
Higgs singlet  superfields are usually  present in most  extensions of
the  Minimal Supersymmetric  Standard  Model (MSSM)  that address  the
$\mu$-problem,  such as  the  Next-to-Minimal Supersymmetric  Standard
Model ({\color{Blue}NMSSM}) and  the Minimal Nonminimal Supersymmetric
Standard Model ({\color{Red}MNSSM}).   Employing a gauge- and flavour-
covariant  effective Lagrangian  formalism,  we show  how the  singlet
Higgs bosons  of such theories  can have significant  contributions to
$B$-meson  flavour-changing  neutral  current (FCNC)  observables  for
large  values of  $\tan\beta \stackrel{>}{{}_\sim}  50$ at  the 1-loop
level.  Illustrative  results are  presented including effects  on the
$B_s$   and   $B_d$  mass   differences   and   on   the  rare   decay
$B_s\to\mu^+\mu^-$.   In particular,  we  find that  depending on  the
actual  value  of  the  lightest  singlet  pseudoscalar  mass  in  the
{\color{Blue}NMSSM}, the branching ratio for $B_s\to\mu^+\mu^-$ can be
enhanced  or  even  suppressed  with  respect to  the  Standard  Model
prediction by more than one order of magnitude.

\newpage


\section{Introduction}

Supersymmetry  (SUSY), softly broken  at the  electroweak (EW)  or TeV
scale,  provides  one  of  the most  self-consistent  frameworks  that
enables  one  to technically  address  the  so-called gauge  hierarchy
problem.   In particular,  SUSY  protects the  EW  scale from  quantum
corrections  that could  be induced  by possible  new dynamics  at the
Grand  Unified  Theory~(GUT)  or   Planck  scale.   Whilst  the  exact
mechanism  of SUSY  breaking  remains unknown  so far,  supersymmetric
theories do in general contain many sources of FCNC's amongst the soft
SUSY-breaking terms.  In order to maintain agreement with experimental
data, the hypothesis of  Minimal Flavour Violation (MFV) is frequently
assumed,   where  the  well-known   Glashow--Iliopoulos--Maiani  (GIM)
mechanism  can be  naturally implemented~\cite{GIM}.   Within  the MFV
framework, all  phenomena of  flavour and CP  violation in  the theory
originate from the Cabbibo-Kobayashi-Maskawa (CKM) mixing matrix ${\bf
V}$ \cite{CKM},  such that all  FCNC interactions vanish in  the limit
${\bf V}\to {\bf 1}_3$.

The FCNC effects  on $B$- and $K$-meson observables  mediated by Higgs
bosons     have     been      well     studied     in     the     MSSM
\cite{MSSMone,MSSMtwo,QCDfactorone,FlavourObservables,MCPMFV}.       At
large values  of $\tan\beta=v_2/v_1$, the  ratio of the  Higgs doublet
vacuum expectation values (VEVs), the couplings of the Higgs bosons to
down-type  quarks becomes  enhanced and  1-loop threshold  effects can
make a significant contribution to  FCNC processes, such as the $B_s -
\bar  B_s$ and  $B_d  - \bar  B_d$  mixings, and  the leptonic  decays
$B_{s,d}\to\mu^+\mu^-$.

On  the other  hand, along  with the  supersymmetric  trilinear Yukawa
interactions, the superpotential of  the MSSM contains a bilinear SUSY
term,  $\mu \hat  H_1 \hat  H_2$, that  couples the  two Higgs-doublet
superfields.  This $\mu$-term is phenomenologically necessary in order
to break the chiral Peccei-Quinn  (PQ) symmetry $U(1)_{\rm PQ}$ and to
provide  a mass  for the  as-yet experimentally  unobserved higgsinos.
The $\mu$-parameter  has dimensions of  mass and is singlet  under the
Standard  Model (SM)  gauge  group.   It can  therefore  be driven  by
supergravity  quantum effects  to  the mass  scales  $M_{\rm GUT}$  or
$M_{\rm Planck}$.   However, successful electroweak  symmetry breaking
in the MSSM implies that $\mu$ must be small, close to $M_{\rm SUSY}$,
namely at the scale at  which SUSY is softly broken.  This theoretical
difficulty  is often  referred to  as  the $\mu$-problem  of the  MSSM
\cite{MUproblem,CPNSH}.

The  $\mu$-problem may  be  resolved by  allowing  an effective  $\mu$
parameter to  be generated at the  SUSY breaking scale, as  the VEV of
(the  scalar   component  of)   an  additional  gauge   singlet  Higgs
superfield,   $\hat  S$.  The   Higgs  bilinear   term  of   the  MSSM
superpotential is replaced with the trilinear coupling $\lambda \hat S
\hat H_1 \hat  H_2$. In its simplest form,  such a model re-introduces
the  undesirable  $U(1)_{\rm  PQ}$  symmetry  of  the  superpotential,
leading to  experimentally excluded electroweak-scale  axions. Several
mechanisms  for  breaking  the  $U(1)_{\rm  PQ}$  symmetry  have  been
discussed    in   the    literature   \cite{CPNSH}    and    lead   to
phenomenologically  distinct  models,   such  as  the  Next-to-Minimal
Supersymmetric  Standard  Model ({\color{Blue}NMSSM})\cite{NMSSM}  and
the     Minimal    Non-minimal    Supersymmetric     Standard    Model
({\color{Red}MNSSM}) \cite{MNSSM}.

An   interesting   limit   for   the   {\color{Blue}NMSSM}   and   the
{\color{Red}MNSSM} is the so-called PQ symmetric limit.  If $U(1)_{\rm
PQ}$ is weakly  broken, then the lightest CP-odd  Higgs field $A_1$ is
predominantly  singlet  and   very  light,  essentially  becoming  the
pseudo-Goldstone  boson  of  this  symmetry.   In  this  case,  it  is
therefore natural to expect that  a pseudoscalar particle $A_1$ with a
mass of  a few GeV  may have eluded  detection so far.   Production of
such  light, highly singlet  pseudoscalars within  the context  of the
{\color{Blue}NMSSM} has been considered  through the decays of SM-like
Higgs  fields  \cite{HtoAAsignals},   in  associated  production  with
charginos  \cite{AssociatedChargino}  and in  rare  decays of  Upsilon
mesons \cite{UpsilonA1}.

In  the {\color{Blue}NMSSM}  and the  {\color{Red}MNSSM},  the singlet
Higgs fields have  no tree-level couplings to the  Standard Model (SM)
fermions or gauge bosons. It has long been known that 1-loop threshold
corrections can  produce significant non-holomorphic  Yukawa couplings
in  the MSSM  at large  $\tan\beta$  \cite{TCone,TCtwo,TCthree}.  Most
recently,   it  has   been   realized~\cite{HodgkinsonPilaftsis}  that
analogous  threshold  corrections  produce sizeable  radiative  Yukawa
couplings even for  the singlet Higgs bosons in  minimal extensions of
the MSSM.

In this paper we consider the  effects of FCNC Yukawa couplings of the
singlet Higgs bosons within the  MFV framework. Such effects have been
considered in \cite{Hiller,Heng} within the {\color{Blue}NMSSM} in the
limit of  a light pseudoscalar singlet.   The scope of  our paper goes
well  beyond  these  studies,  both  analytically,  as  our  effective
Lagrangian  approach includes  a resummation  of  $\tan\beta$ enhanced
terms, but also phenomenologically,  as we discuss the implications of
light  singlets  at  or  below  the  electroweak  scale  in  both  the
{\color{Blue}NMSSM} and the {\color{Red}MNSSM}.

The structure  of our paper is as  follows: in Section 2  we present a
manifestly gauge- and flavour-covariant effective Lagrangian framework
for  the calculation  of FCNC  vertices between  the Higgs  bosons and
fermions.   Section   3  summarises  the   relevant  analytic  results
regarding  FCNC $B$-meson observables.  Numerical results  for various
FCNC processes are  presented in Section 4, for  relevant scenarios in
both   the  {\color{Red}MNSSM}   and  the   {\color{Blue}NMSSM}.   Our
conclusions are given in Section 5.


\section{Effective Lagrangian Formalism}
\label{EffectiveLagrangian}
\setcounter{equation}{0}

\begin{figure}[t!]
\begin{center}
\includegraphics[scale=0.7]{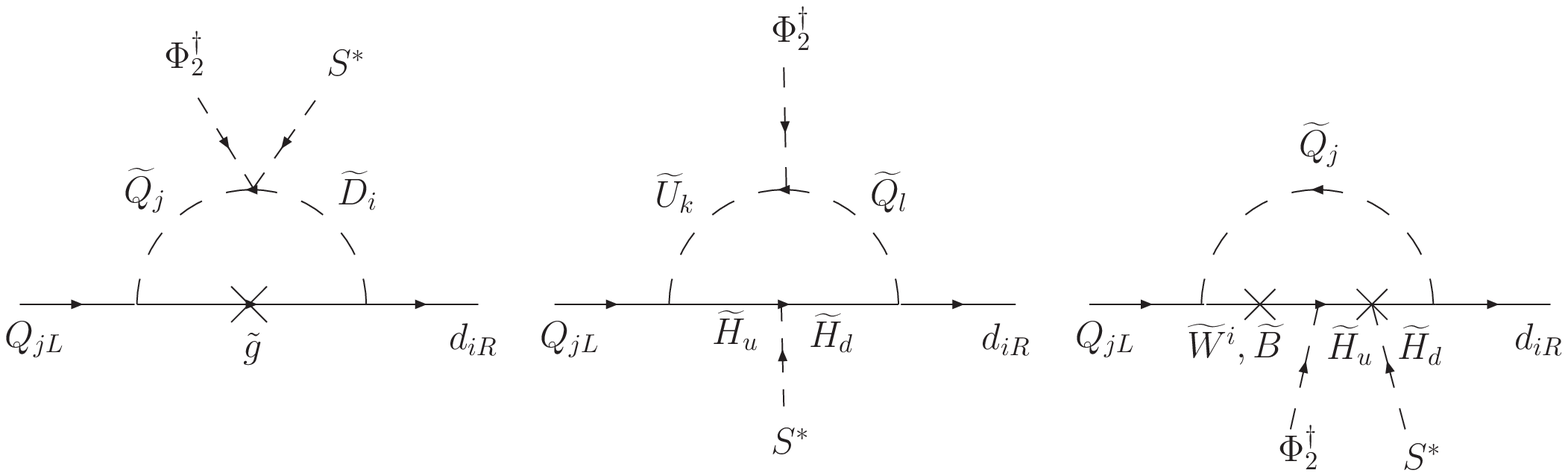}
\caption{\it Gauge-  and flavour-invariant self-energy  graphs for the
down-type quarks  to leading  order in the  Higgs fields  $\Phi_2$ and
$S$.}  \label{SelfEn}
\end{center}
\end{figure}

In  this section  we  derive the  form  of the  manifestly gauge-  and
flavour-covariant  effective Lagrangian  for Higgs  boson interactions
with  fermions.  Our  conventions  and notations  here closely  follow
those  of  \cite{MCPMFV}.  The  effective  Lagrangian  describing  the
down-type quark  self-energy transition $Q^0_{jL}\to  d^0_{iR}$ may be
written in a gauge-symmetric and flavour-covariant form as
\begin{equation}
  \label{SELagrangian}
-{\mathcal L}_{\rm eff}^d [\Phi_1,\Phi_2,S]
=
\bar{d^0_{iR}}\left({\bf h}_d \Phi_1^\dag
+\Delta{\bf h}_d [\Phi_1,\Phi_2,S]\right)_{ij}
Q^0_{jL}\ +\ {\rm H.c.}\ ,
\end{equation}
where  $\Phi_{1(2)}$ are the  scalar components  of the  Higgs doublet
superfields   giving  masses   to  the   down-type   (up-type)  quarks
respectively\footnote{Here  we  adopt  the  convention for  the  Higgs
doublets:  $H_u\equiv\Phi_2,\  H_d\equiv  i\tau_2 \Phi_1^\ast$,  where
$\tau_2$ is the  usual Pauli matrix.}, $S$ is  the scalar component of
the  Higgs singlet  superfield  and  ${\bf h}_d$  is  the $3\times  3$
down-type  Yukawa coupling  matrix. In~(\ref{SELagrangian})  the first
term is  the tree  level contribution, whilst  $\Delta{\bf h}_d$  is a
$3\times   3$   matrix   which   is  a   Coleman--Weinberg   effective
functional\cite{ColemanWeinberg}   of  the  background   Higgs  fields
$\Phi_{1,2}$  and $S$.  Observe that  the 1-loop  effective functional
$\Delta{\bf  h}_d$  has  the  same flavour-  and  gauge-transformation
properties  as ${\bf h}_d\,  \Phi^\dagger_1$.  Typical  Feynman graphs
that contribute  to $\Delta{\bf  h}_d$ to leading  order in  the Higgs
fields  $\Phi_2$  and  $S$  are displayed  in  Fig.~\ref{SelfEn}.  In
detail, the analytic form of $\Delta{\bf h}_d$ is
\begin{eqnarray}
  \label{matrixintegral}
\left(\Delta{\bf h}_d\right)_{ij} & = & 
\int {d^n k\over(2\pi)^n i}
\left[P_L {2C_F\,g_3^2 M_3^\ast\over k^2-|M_3^2|}
\left({1\over k^2{\bf 1}_{12}-\widetilde{\bf M}^2}\right)_
{\tilde D_i\tilde Q_j^\dag}\right.\nonumber\\
&&\hspace{-2cm} +\ P_L\left({1\over \slash k {\bf 1}_9-{\bf M}_C P_L
-{\bf M}_C^\dag P_R}\right)_{\tilde H_d \tilde H_u}
P_L \left({\bf h}_d\right)_{il}
\left({1\over k^2{\bf 1}_{12}-\widetilde{\bf M}^2}\right)_
{\tilde Q_l \tilde U_k^\dag} \left({\bf h}_u\right)_{kj}\nonumber\\
&&\hspace{-2cm} + P_L 
\left({1\over \slash k{\bf 1}_9
-{\bf M}_C P_L-{\bf M}_C^\dag P_R}\right)_{\tilde H_d \tilde B}P_L
\left({\bf h}_d\right)_{il}
\left({1\over k^2{\bf 1}_{12}-\widetilde{\bf M}^2}\right)_{\tilde Q_l
\tilde Q_j^\dag}(\sqrt{2} g_1)\\
&&\hspace{-2cm} +\sum_{k=1}^3 \left. P_L 
\left({1\over \slash k{\bf 1}_9
-{\bf M}_C P_L-{\bf M}_C^\dag P_R}\right)_{\tilde H_d \tilde W^k}P_L
\left({\bf h}_d\right)_{il}
\left({1\over k^2{\bf 1}_{12}-\widetilde{\bf M}^2}\right)_{\tilde Q_l
\tilde Q_j^\dag}\ \left(\frac{g_2\tau^k}{\sqrt{2}}\right)
 \right]\ .\nonumber
\end{eqnarray}
Here  $n=4-2\epsilon$  is  the  number of  dimensions  in  dimensional
regularisation  (DR),  ${\bf  1}_N$  is  the  $N\times  N$-dimensional
identity     matrix,    $\tau^{1,2,3}$ are  the     Pauli    matrices,
$P_{L(R)}=\frac{1}{2}[1-(+)\gamma_5]$   are  the   standard  chirality
projection operators  and $C_F$ is the quadratic  Casimir invariant of
QCD in the fundamental representation, i.e.~$C_F = 4/3$.  Moreover, we
denote  the  $U(1)_Y$,  $SU(2)_L$  and $SU(3)_c$  gauge  couplings  by
$g_{1,2,3}$ and the soft SUSY-breaking gluino mass by~$M_3$.  Finally,
$\widetilde{\bf M}^2$ and ${\bf  M}_C$ are $12\times 12$- and $9\times
9$-dimensional    matrices    which    describe   the    squark    and
chargino-neutralino mass matrices,  respectively, in the background of
non-vanishing Higgs fields $\Phi_{1,2}$ and $S$.

We express the $9\times 9$ chargino-neutralino mass matrix ${\bf M}_C$
in  the  Weyl basis  $$\left(\widetilde  B,\, \widetilde  W^{1,2,3},\,
\widetilde   H_u,\,  \widetilde   H_d,\,   \tilde  s\right),$$   where
$\widetilde  H_{u,d}$   are  $SU(2)_L$  doublets   $\widetilde  H_u  =
\left(\tilde   h_u^+,  \tilde   h_u^0\right)^T$,  $\widetilde   H_d  =
\left(\tilde  h_d^0,  \tilde h_d^-\right)^T$  and  $\tilde  s$ is  the
singlino  Higgs field.   In  the {\color{Blue}NMSSM},  ${\bf M}_C$  is
given by
\begin{equation}
\label{charginomassmatrix}
{\bf M}_C [\Phi_1,\Phi_2,S] =
\left(\begin{array} {ccccc}
M_1 & 0 & -\frac{1}{\sqrt 2} g_1 \Phi_2^\dag &
\frac{1}{\sqrt 2} g_1\Phi_1^T (i\tau_2) & 0\\
0 & M_2{\bf 1}_3 & \frac{1}{\sqrt 2} g_2 \Phi_2^\dag \tau_i
& -\frac{1}{\sqrt 2} g_2 \Phi_1^T (i\tau_2) \tau_i & 0\\
-\frac{1}{\sqrt 2} g_1 \Phi_2^\ast & \frac{1}{\sqrt 2} g_2
\tau_i^T\Phi_2^\ast & {\bf 0}_2 & \lambda S(i\tau_2) & -\lambda\Phi_1^\ast\\
-\frac{1}{\sqrt 2}(i\tau_2) g_1 \Phi_1 & \frac{1}{\sqrt 2} g_2 \tau_i^T
(i\tau_2)\Phi_1 & -\lambda S(i\tau_2) & {\bf 0}_2 &
-\lambda(i\tau_2)\Phi_2\\
0 & 0 & -\lambda\Phi_1^\dag & \lambda\Phi_2^T(i\tau_2) & 2\kappa S
\end{array} \right)\ ,
\end{equation}
where $M_{1,2}$  are the bino  and wino soft SUSY-breaking  masses and
$\kappa$ is the trilinear  singlet coupling of the {\color{Blue}NMSSM}
superpotential; the corresponding matrix for the {\color{Red}MNSSM} is
obtained by  setting $\kappa=0$. The $12\times 12$  squark mass matrix
$\widetilde{\bf  M}^2$  is  identical   in  both  models.    
More explicitly, it is given by
\begin{equation}
\label{squarkmassmatrix}
\widetilde{\bf M}^2 [\Phi_1,\Phi_2,S]=
\left(\begin{array} {ccc}
\left(\widetilde{\bf M}^2\right)_{\tilde Q^\dag \tilde Q} &
\left(\widetilde{\bf M}^2\right)_{\tilde Q^\dag \tilde U} &
\left(\widetilde{\bf M}^2\right)_{\tilde Q^\dag \tilde D} \\ 
\left(\widetilde{\bf M}^2\right)_{\tilde U^\dag \tilde Q} &
\left(\widetilde{\bf M}^2\right)_{\tilde U^\dag \tilde U} &
\left(\widetilde{\bf M}^2\right)_{\tilde U^\dag \tilde D} \\
\left(\widetilde{\bf M}^2\right)_{\tilde D^\dag \tilde Q} &
\left(\widetilde{\bf M}^2\right)_{\tilde D^\dag \tilde U} &
\left(\widetilde{\bf M}^2\right)_{\tilde D^\dag \tilde D}
\end{array}\right)_{ij}\ \ ,
\end{equation}
with
\begin{eqnarray}
\label{squarkmatrixelements}
\left(\widetilde{\bf M}^2\right)_{\tilde Q_i^\dag \tilde Q_j} & = &
\left(\widetilde{\bf M}_Q^2\right)_{ij}{\bf 1}_2
+\left({\bf h}^\dag_d{\bf h}_d\right)_{ij}\Phi_1\Phi_1^\dag
+\left({\bf h}^\dag_u{\bf h}_u\right)_{ij}
\left(\Phi_2^\dag\Phi_2{\bf 1}_2-\Phi_2\Phi_2^\dag\right)\nonumber\\
&&
-\frac{1}{2}\delta_{ij} g_2^2\left(\Phi_1\Phi_1^\dag-\Phi_2\Phi_2^\dag\right)
+\delta_{ij}\left(\frac{1}{4}g_2^2-\frac{1}{12}{g_1}^2\right)
\left(\Phi_1^\dag\Phi_1-\Phi_2^\dag\Phi_2\right){\bf 1}_2,\nonumber\\
\left(\widetilde{\bf M}^2\right)_{\tilde U^\dag_i\tilde Q_j} & = &
\left(\widetilde{\bf M}^2\right)^\dag_{\tilde Q^\dag_j\tilde U_i}  =
-\left({\bf a}_u\right)_{ij}\Phi_2^T i\tau_2
+\left({\bf h}_u\right)_{ij}\lambda S^\ast \Phi_1^T i\tau_2 ,\nonumber\\
\left(\widetilde{\bf M}^2\right)_{\tilde D^\dag_i\tilde Q_j} & = &
\left(\widetilde{\bf M}^2\right)^\dag_{\tilde Q^\dag_j\tilde D_i}  =
+\left({\bf a}_d\right)_{ij}\Phi_1^\dag
-\left({\bf h}_d\right)_{ij}\lambda S^\ast \Phi_2^\dag ,\nonumber\\
\left(\widetilde{\bf M}^2\right)_{\tilde U_i^\dag \tilde U_j} & = &
\left(\widetilde{\bf M}^2_U\right)_{ij}
+\left({\bf h}_u{\bf h}_u^\dag\right)_{ij}\Phi_2^\dag\Phi_2
+\frac{1}{3}\delta_{ij} {g_1}^2\left(\Phi_1^\dag\Phi_1-\Phi_2^\dag\Phi_2
\right) ,\nonumber\\
\left(\widetilde{\bf M}^2\right)_{\tilde D_i^\dag \tilde D_j} & = &
\left(\widetilde{\bf M}^2_D\right)_{ij}
+\left({\bf h}_d{\bf h}_d^\dag\right)_{ij}\Phi_1^\dag\Phi_1
-\frac{1}{6}\delta_{ij} {g_1}^2\left(\Phi_1^\dag\Phi_1-\Phi_2^\dag\Phi_2
\right) ,\nonumber\\
\left(\widetilde{\bf M}^2\right)_{\tilde U_i^\dag \tilde D_j} & = &
\left(\widetilde{\bf M}^2\right)^\dag_{\tilde D_j^\dag \tilde U_i} =
\left({\bf h}_u{\bf h}_d^\dag\right)_{ij}\Phi_1^T i\tau\Phi_2\ .
\end{eqnarray}
In the above,  ${\bf h}_u$ is the $3\times  3$ up-type Yukawa coupling
matrix,  $\widetilde{\bf  M}^2_{Q,U,D}$   are  the  $3\times  3$  soft
mass-squared   matrices  of  the   squarks  and   ${\bf  a}_{d,u}={\bf
h}_{d,u}{\bf A}_{d,u}$  are the corresponding $3\times  3$ soft Yukawa
matrices.   Notice that (\ref{squarkmassmatrix})  reduces to  the MSSM
result  presented in~\cite{MCPMFV},  after  identifying the  effective
$\mu$-parameter  as $\mu =  - \frac{1}{\sqrt{2}}\,\lambda  v_S$, where
$v_S = \sqrt{2}\,\langle S\rangle$ is the VEV of $S$.

The  weak  left-  and  right-handed quarks,  denoted  as  $u^0_{L,R}$,
$d^0_{L,R}$, are related to  the respective mass eigenstates $u_{L,R}$
and $d_{L,R}$ by the unitary transformations
\begin{equation}
\label{weakmasstransform}
u^0_L = {\bf U}^Q_L \ u_L,\ \ \ 
d^0_L = {\bf U}^Q_L\  {\bf V} \ d_L,\ \ \ 
u^0_R = {\bf U}^u_R \ u_R,\ \ \ 
d^0_R = {\bf U}^d_R \ d_R.
\end{equation}
Here  ${\bf U}^Q_L$  and  ${\bf U}^{u,d}_R$  are  unitary $3\times  3$
matrices,  and  ${\bf V}$  is  the  physical  CKM-mixing matrix.   The
matrices  ${\bf U}^Q_L$ and  ${\bf U}^{u,d}_R$  are determined  by the
mass renormalisation conditions
\begin{equation}
\label{quarkmassconditions}
\left<{\mathcal L}^d_{\rm eff} \left[\Phi_1,\Phi_2,S\right]\right>
\ =\ - \bar{d}_R \widehat {\bf M}_d d_L\ +\ {\rm H.c.},\ \ \ 
\left<{\mathcal L}^u_{\rm eff} \left[\Phi_1,\Phi_2,S\right]\right>
\ =\ - \bar{u}_R \widehat {\bf M}_u u_L\ +\ {\rm H.c.},
\end{equation}
where $\left<\ldots\right>$ indicates  the vacuum expectation value of
the   enclosed   expression.   In~(\ref{quarkmassconditions}),   ${\bf
M}_{d,u}$  are  the physical  down-  and  up-quark  mass matrices  and
${\mathcal L}^u_{\rm  eff}$ is the  corresponding effective Lagrangian
for  up-type Yukawa  sector [cf.~(\ref{SELagrangian})].   Imposing the
conditions~(\ref{quarkmassconditions}) yields
\begin{equation}
\label{yukawaconditions}
{\bf U}_R^{d \dag}\  {\bf h}_d\  {\bf U}^Q_L
\ =\
\frac{\sqrt{2}}{v_1} \widehat{\bf M}_d\  {\bf V}^\dag
\  {\bf R}_d^{-1},\ \ \ 
{\bf U}_R^{u \dag}\  {\bf h}_u\  {\bf U}^Q_L
\ =\
\frac{\sqrt{2}}{v_2} \widehat{\bf M}_u\  {\bf R}_u^{-1},\ \ \ 
\end{equation}
where
\begin{eqnarray}
{\bf R}_d  \!& = &\! {\bf 1}_3\ +\ \frac{\sqrt 2}{v_1} {\bf U}^{Q\dag}_L
\left<{\bf h}_d^{-1} \Delta{\bf h}_d\left[\Phi_1,\Phi_2,S\right]
\right>{\bf U}_L^Q\ ,\nonumber\\
{\bf R}_u \!& = &\! {\bf 1}_3\ +\ \frac{\sqrt 2}{v_2} {\bf U}^{Q\dag}_L
\left<{\bf h}_u^{-1} \Delta{\bf h}_u\left[\Phi_1,\Phi_2,S\right]
\right>{\bf U}_L^Q\ .
\end{eqnarray}
Note  that  $\Delta{\bf  h}_u$  is  the  respective  1-loop  effective
functional for the up-type quark Yukawa couplings.

Considering       now      the       general       effective      FCNC
Lagrangian~(\ref{SELagrangian})  for Higgs  interactions  to down-type
quarks, we find that
\begin{eqnarray}
\label{EffIntLagrangian}
-{\mathcal L}_{\rm FCNC}^{d,H} & = & 
\bar d_R \frac{{\bf h}_d}{\sqrt{2}}
\left[\phi_1\left({\bf 1}_3+{\bf \Delta}_d^{\phi_1}\right)
-i a_1\left({\bf 1}_3+{\bf \Delta}_d^{a_1}\right)\right.
\nonumber\\
&&\left.\ \ \ 
+\phi_2{\bf \Delta}_d^{\phi_2}-i a_2 {\bf \Delta}_d^{a_2}
+\phi_S{\bf \Delta}_d^{\phi_S}-i a_S {\bf \Delta}_d^{a_S}\right]
{\bf V} d_L
\nonumber\\
&&
+\bar d_R {\bf h}_d \left[\phi_1^-\left({\bf 1}_3+{\bf \Delta}_d^{\phi_1^-}
\right)+\phi_2^-{\bf \Delta}_d^{\phi_2^-}\right]u_L\ +\ {\rm H.c.},
\end{eqnarray}
where the component Higgs fields are given by
\begin{equation}
\label{componentfields}
\Phi_{1,2}\ =\ \left(\begin{array}{c}
\phi_{1,2}^+\\
\frac{1}{\sqrt{2}}\left(v_{1,2}+\phi_{1,2}+i a_{1,2}\right)
\end{array}\right),\qquad 
S\ =\ \frac{1}{\sqrt{2}}\left(v_S+\phi_S+i a_S\right).
\end{equation}
In~(\ref{EffIntLagrangian}),   ${\bf   \Delta}_d^{\phi_{1,2,S}},  {\bf
\Delta}_d^{a_{1,2,S}}$   and  ${\bf   \Delta}_d^{\phi^\pm_{1,2}}$  are
3-by-3 matrices that are evaluated as
\begin{equation}
\label{hlet}
{\bf \Delta}_d^{\phi_{1,2,S}} = \sqrt{2}
\left<\frac{\delta}{\delta\phi_{1,2,S}} {\bf \Delta}_d\right>,\ \ \ 
{\bf \Delta}_d^{a_{1,2,S}} = i\sqrt{2}
\left<\frac{\delta}{\delta a_{1,2,S}} {\bf \Delta}_d\right>,\ \ \ 
{\bf \Delta}_d^{\phi^\pm_{1,2}} = 
\left<\frac{\delta}{\delta\phi^\pm_{1,2}} {\bf \Delta}_d\right>,
\end{equation}
where we have defined ${\bf \Delta}_d \equiv {\bf h}_d^{-1} \Delta{\bf
h}_d[\Phi_1,\Phi_2,S]$   and  suppressed  the   vanishing  iso-doublet
components on  the LHS of~(\ref{hlet}). Since the  singlet Higgs field
does not  contain a  charged component, the  couplings of  the charged
Higgs boson  are unchanged  from those  of the MSSM  and we  shall not
consider them here  any further.  In the neutral  sector, the physical
Higgs boson  mass eigenstates are  related to the weak  eigenstates by
the orthogonal mixing matrices $O^H$ and $O^A$, such that
\begin{equation}
\label{mixingmatrices}
\left(\begin{array}{c}
\phi_1\\
\phi_2\\
\phi_S\end{array}\right)\
=\ O^H \left(\begin{array}{c}
H_1\\
H_2\\
H_3\end{array}\right),\qquad \left(\begin{array}{c}
a\\
a_S\end{array}\right)\ =\ O^A \left(\begin{array}{c}
A_1\\
A_2\end{array}\right)\ .
\end{equation}
It is sometimes useful to parameterise $O^A$ by the CP-odd mixing
angle $\theta_A$, where
\begin{equation}
O^A = \left(\begin{array}{cc}
\cos\theta_A & \sin\theta_A \\
-\sin\theta_A & \cos\theta_A
\end{array}\right)\ .
\end{equation}

In   terms   of   the   mass  eigenstates,   the   general   effective
Lagrangian~(\ref{SELagrangian}) that includes the FCNC interactions of
the neutral Higgs bosons with the down-type quarks reads:
\begin{eqnarray}
-{\mathcal L}^{d,H}_{\rm FCNC} & = &
\frac{g}{2M_W}\left[H_i \bar d_R
\left(\widehat{\bf M}_d\ {\bf g}^L_{H_i \bar dd}P_L
+{\bf g}^R_{H_i \bar dd}\ \widehat{\bf M}_d P_R\right) d\right.
\nonumber\\
&&\left.+\ A_j \bar d_R
\left(\widehat{\bf M}_d\ {\bf g}^L_{A_j \bar dd}P_L
+{\bf g}^R_{A_j \bar dd}\ \widehat{\bf M}_d P_R\right) d\right]\; ,
\end{eqnarray}
where the Higgs couplings in the flavour basis
${\bf U}^Q_L={\bf U}^u_R={\bf U}^d_R={\bf 1}_3$ are given by
\begin{eqnarray}
\label{CouplingMatrices}
{\bf g}^L_{H_i \bar dd} & = &
\frac{{\mathcal O}^H_{1i}}{c_\beta}{\bf V}^\dag
{\bf R}_d^{-1}\left({\bf 1}_3+{\bf \Delta}_d^{\phi_1}\right)
{\bf V}
+\frac{{\mathcal O}^H_{2i}}{c_\beta}{\bf V}^\dag
{\bf R}_d^{-1}{\bf \Delta}_d^{\phi_2}{\bf V}\nonumber\\
&& +\frac{{\mathcal O}^H_{3i}}{c_\beta}{\bf V}^\dag
{\bf R}_d^{-1}{\bf \Delta}_d^{\phi_S}{\bf V}\ ,\\
{\bf g}^R_{H_i \bar dd} & = & 
\left({\bf g}^L_{H_i \bar dd}\right)^\dag\ ,\\
{\bf g}^L_{A_i \bar dd} & = &
i {\mathcal O}^A_{1i} t_\beta {\bf V}^\dag {\bf R}_d^{-1}
\left({\bf 1}_3+{\bf \Delta}_d^{a_1}-\frac{1}{t_\beta}
{\bf \Delta}_d^{a_2}\right){\bf V}\nonumber\\
&&-i \frac{{\mathcal O}^{A}_{2i}}{c_\beta}{\bf V}^\dag
{\bf R}_d^{-1}{\bf \Delta}_d^{a_S}{\bf V}\ ,\\
{\bf g}^R_{A_i \bar dd} & = & 
\left({\bf g}^L_{A_i \bar dd}\right)^\dag\ .
\end{eqnarray}
In     the    above,    we     used    the     short-hand    notation:
$t_\beta\equiv\tan\beta=v_2/v_1$,     $c_\beta\equiv\cos\beta$     and
$s_\beta\equiv\sin\beta$.

\pagebreak

\subsection{Single Higgs Insertion Approximation}

Our numerical results of Section~\ref{NumericalResults} include a full
evaluation of  ${\bf \Delta}_d$ and  its derivatives using  the matrix
expression of~(\ref{matrixintegral}).  We  may obtain an understanding
of our  results by considering an  expansion of $\Delta  {\bf h}_d$ in
terms  of $\Phi_2$  and  $S$. We  call  the leading  term  of such  an
expansion,     as     represented     by    Fig.~\ref{SelfEn},     the
single-Higgs-insertion (SHI) approximation.

Within  the  SHI  approximation,  the  $\tan\beta$-enhanced  threshold
corrections satisfy the simple relation
\begin{equation}
\frac{\sqrt{2}}{v_2}\left<{\bf \Delta}_d\right>
={\bf \Delta}_d^{\phi_2}
={\bf \Delta}_d^{a_2}
=\frac{v_S}{v_2}{\bf \Delta}_d^{\phi_S}
=\frac{v_S}{v_2}{\bf \Delta}_d^{a_S}\ ,
\end{equation}
where
\begin{equation}
\frac{\sqrt{2}}{v_2}\left<{\bf \Delta}_d\right>\
=\
{\bf 1}_3\frac{2\alpha_S}{3\pi}\mu M_3 I(\widetilde M_Q^2,\widetilde
M_D^2,M_3^2)\: +\:
\frac{{\bf h}_u^\dag{\bf h}_u}{16\pi^2}\mu A_u I(\widetilde
M_Q^2,\widetilde M_U^2,\mu^2)\ , 
\end{equation}
and $I(x,y,z)$ is the 1-loop function,
\begin{equation}
I(x,y,z)=
\frac{xy\ln(x/y)+yz\ln(y/z)+xz\ln(z/x)}
{(x-y)(y-z)(x-z)}\ .
\end{equation}
In writing the above,  we have neglected the subdominant contributions
coming  from  chargino-  and  neutralino-exchange  graphs,  which  are
weak-coupling  constant  suppressed.

%
\section{$B$-meson FCNC Observables}
\label{BPhysics}
\setcounter{equation}{0}

In  this section  we  review  the analytic  results  relevant to  FCNC
$B$-meson   observables.   Our   conventions  and   discussion  follow
\cite{FlavourObservables}.

\subsection{$\Delta M_{B_q}$}

In the approximation of equal $B$-meson lifetimes, the SM and SUSY
contributions may be written separately as
\begin{equation}
\Delta M_{B_q} = 2 \left|\left< \bar B^0_q \right|
H_{\rm eff}^{\Delta B=2} \left| B^0_q \right>_{\rm SM} +
\left< \bar B^0_q \right|
H_{\rm eff}^{\Delta B=2} \left| B^0_q \right>_{\rm SUSY}\right| \ .
\end{equation}
Although the SM predictions are consistent with the observed experimental
values \cite{LN}, the uncertainties are large and a non-negligible SUSY contribution
is not excluded.

\vfill\eject

The SUSY contributions are given by
\begin{eqnarray}
\label{MixingHamiltonian}
\left< \bar B^0_d \right|H_{\rm eff}^{\Delta B=2}
\left| B^0_d \right>_{\rm SUSY}
& = & 1711 {\rm ps}^{-1} \left(\frac{\hat B^{1/2}_{B_d} F_{B_d}}
{230\  {\rm MeV}}\right)^2 \left(\frac{\eta_B}{0.55}\right)\nonumber \\
&&
\hspace{-1.5cm}
\!\!\!\!\times\left[0.88\left(C_2^{\rm LR(DP)}+C_2^{\rm LR(2HDM)}\right)
-0.52\left(C_1^{\rm SLL(DP)}+C_1^{\rm SRR(DP)}\right)\right]\ ,\\
\left< \bar B^0_s \right|H_{\rm eff}^{\Delta B=2}
\left| B^0_s \right>_{\rm SUSY}
& = & 2310 {\rm ps}^{-1} \left(\frac{\hat B^{1/2}_{B_s} F_{B_s}}
{265\  {\rm MeV}}\right)^2 \left(\frac{\eta_B}{0.55}\right)\nonumber \\
&&
\hspace{-1.5cm}
\!\!\!\!\times\left[0.88\left(C_2^{\rm LR(DP)}+C_2^{\rm LR(2HDM)}\right)
-0.52\left(C_1^{\rm SLL(DP)}+C_1^{\rm SRR(DP)}\right)\right]\ ,
\end{eqnarray}
where DP  stands for the  Higgs-mediated double-penguin contributions.
Here $F_{B_q}$ is the weak decay constant and $B_{B_q}$ is the
so-called ``Bag'' parameter of the $B^0_q$ meson respectively.
The numerical factor $\eta_B$ is due to QCD corrections.  We use the
next-to-leading order QCD factors \cite{QCDfactorone,QCDfactortwo} and
hadronic matrix elements  at the scale $\mu=4.2$ GeV:
\begin{equation}
\bar P_1^{\rm LR} = -0.58\ ,
\hspace{1cm}
\bar P_2^{\rm LR} = 0.88\ ,
\hspace{1cm}
\bar P_1^{\rm SLL} = -0.52\ ,
\hspace{1cm}
\bar P_2^{\rm SLL} = -1.1\ .
\end{equation}

The double-penguin Wilson coefficients in~(\ref{MixingHamiltonian}) are given by
\begin{eqnarray}
C_1^{\rm SLL(DP)} & = & \ 
\frac{-16\pi^2 m_b^2}{\sqrt{2}G_F M_W^2}\ 
\left(\sum_{i=1}^3 
\frac{{\bf g}^L_{H_i \bar bq} {\bf g}^L_{H_i \bar bq}}
{M^2_{H_i}}\
+\ \sum_{j=1}^2 
\frac{{\bf g}^L_{A_j \bar bq} {\bf g}^L_{A_j \bar bq}}
{M^2_{A_j}}\right)\; ,
\nonumber\\
C_1^{\rm SRR(DP)} & = & \ 
\frac{-16\pi^2 m_q^2}{\sqrt{2}G_F M_W^2}\ 
\left(\sum_{i=1}^3 
\frac{{\bf g}^R_{H_i \bar bq} {\bf g}^R_{H_i \bar bq}}
{M^2_{H_i}}\
+\ \sum_{j=1}^2 
\frac{{\bf g}^R_{A_j \bar bq} {\bf g}^R_{A_j \bar bq}}
{M^2_{A_j}}\right)\; ,\\
C_2^{\rm LR(DP)} & = &
\frac{-32\pi^2 m_b m_q}{\sqrt{2}G_F M_W^2}
\left(\sum_{i=1}^3 
\frac{{\bf g}^L_{H_i \bar bq} {\bf g}^R_{H_i \bar bq}}
{M^2_{H_i}}\ 
+\ \sum_{j=1}^2 
\frac{{\bf g}^L_{A_j \bar bq} {\bf g}^R_{A_j \bar bq}}
{M^2_{A_j}}\right)\; ,\nonumber
\end{eqnarray}
where the  relevant couplings  ${\bf g}^{L,R}_{H_i(A_j) \bar  bq}$ are
given in~(\ref{CouplingMatrices}). We neglect the $B$-meson masses and
Higgs boson  widths in  the denominators of  the sums, except  for the
lightest Higgs  mass eigenstates  $H_1$ and $A_1$,  since we  may have
masses  ${\mathcal O}(1\ {\rm  GeV})$ in  the case  of a  highly gauge
singlet particle.  In this  limit we replace the effective propagators
$M^{-2}_{H_1(A_1)}$ by their Breit-Wigner  forms.  Of the two relevant
1-loop   contributions    to   $\left<\bar   B^0|H_{\rm   eff}^{\Delta
B=2}|B^0\right>$, the $t-H^\pm$ box contribution to $C_2^{\rm LR}$ may
be given to a good approximation by \cite{QCDfactorone}
\begin{equation}
C_2^{\rm LR(2HDM)} \approx
-\frac{2 m_b m_q}{M_W^2}(V^\ast_{tb}V_{tq})^2 \tan^2\beta\ .
\end{equation}
We also include the chargino-stop box diagram which contributes to
$C_1^{\rm SLL}$, although its contribution remains subdominant.

\subsection{$\bar B^0_{d,s} \to \mu^+\mu^-$}

Neglecting contributions proportional to the lighter quark masses
$m_{d,s}$, the effective Hamiltonian for $\Delta B=1$ FCNC processes
is given by
\begin{equation}
  \label{HeffB1}
H_{\rm eff}^{\Delta B=1}\ =\ -2\sqrt{2} G_F V_{tb} V^\ast_{tq}
\left(C_S {\mathcal O}_S +C_P {\mathcal O}_P+C_{10} {\mathcal O}_{10}
\right)\ ,
\end{equation}
where
\begin{eqnarray}
{\mathcal O}_S & = &
\frac{e^2}{16\pi^2}m_b
\left(\bar q P_R b\right)
\left(\bar\mu\mu\right)\ ,\nonumber\\
{\mathcal O}_P & = &
\frac{e^2}{16\pi^2}m_b
\left(\bar q P_R b\right)
\left(\bar\mu\gamma_5 \mu\right)\ ,\nonumber\\
{\mathcal O}_{10} & = &
\frac{e^2}{16\pi^2}m_b
\left(\bar q\gamma^\mu P_L b\right)
\left(\bar\mu\gamma_\mu\gamma_5 \mu\right)\ .
\end{eqnarray}
The Wilson coefficients $C_S$ and $C_P$ are given at large $\tan\beta$
by
\begin{eqnarray}
C_S & = &
\ \ \frac{2\pi m_\mu}{\alpha_{\rm em}}\frac{1}{V_{tb}V^\ast_{tq}}
\sum_{i=1}^3 \frac{{\bf g}^R_{H_i \bar q b}{\bf g}^S_{H_i\bar\mu\mu}}
{M^2_{H_i}}\ ,
\nonumber\\ 
C_P & = &
i\ \frac{2\pi m_\mu}{\alpha_{\rm em}}\frac{1}{V_{tb}V^\ast_{tq}}
\sum_{i=1}^2 \frac{{\bf g}^R_{A_i \bar q b}{\bf g}^P_{A_i\bar\mu\mu}}
{M^2_{A_i}}\ ,
\end{eqnarray}
and  $C_{10}=-4.221$ denotes  the  leading SM  contribution. Again  we
neglect the  $B$-meson masses  and Higgs boson  widths except  for the
contribution  of  the  lightest  Higgs scalar  and  pseudoscalar.  The
reduced  scalar and  pseudoscalar Higgs  couplings to  charged leptons
${\bf     g}^{S(P)}_{H_i(A_i)\bar    l     l}$     are    given     in
\cite{HodgkinsonPilaftsis}.   The   leptonic  vertex  corrections  can
become important in the limit  of a light, highly singlet Higgs boson,
so we retain them in our numerical estimates.

Using  the  effective $\Delta  B  =1$ Hamiltonian~(\ref{HeffB1}),  the
branching  ratio  for $\bar  B^0_{d,s}  \to  \mu^+\mu^-$  is given  by
\cite{BtoMuMu}
\begin{equation}
\label{BranchingRateEquation}
{\mathcal B}(\bar B^0_{d,s} \to \mu^+\mu^-) = 
\frac{G^2_F\alpha^2_{\rm em}}{16\pi^3}
M_{B_q}\tau_{B_q} |V_{tb}V^\ast_{tq}|^2
\sqrt{1-\frac{4 m_\mu^2}{M_{B_q}^2}}
\left[\left(1-\frac{4 m_\mu^2}{M_{B_q}^2}\right)|F_S^q|^2
+|F_P^q+2m_\mu F_A^q|^2\right],
\end{equation}
where $q=d,s$ and $\tau_{B_q}$ is the total lifetime of the $B_q$
meson. The form factors $F^q_{S,P,A}$ are given by
\begin{equation}
F^q_{S,P} = -\frac{i}{2}M^2_{B_q}F_{B_q}
\frac{m_b}{m_b+m_q} C_{S,P}\ ,
\hspace{1cm}
F_A^q = -\frac{i}{2} F_{B_q} C_{10}\ .
\end{equation}

\pagebreak

\section{Numerical Results}
\setcounter{equation}{0}
\label{NumericalResults}

In this  section we provide  numerical estimates of  FCNC observables,
within   the    contexts   of   both    the   {\color{Red}MNSSM}   and
{\color{Blue}NMSSM}. We focus on those sectors of the models where the
gauge-singlet Higgs fields  are predicted to be light.   We assume the
framework of MFV so that all flavour changing effects are proportional
to the CKM matrix ${\bf  V}$.  This implies that the quark-sector soft
SUSY-breaking  terms are  proportional  to ${\bf  1}_3$~\cite{MCPMFV},
i.e.
\begin{equation}
\widetilde {\bf M}^2_{Q,U,D} = \widetilde M^2_{Q,U,D}\ {\bf 1}_{3},\qquad
{\bf A}_{u,d} = A_{u,d}\ {\bf 1}_3\ ,
\end{equation}
and similarly for the leptonic sector.  We neglect the
neutralino-mediated contributions 
to the FCNC Higgs boson couplings, which are found to be subdominant.

Important constraints on FCNCs come from  $B\to X_s\gamma$ experiments,
which are  sensitive to large stop-quark  mixing due  to a  dominant
chargino-stop  quark  loop.  To avoid conflict  with these constraints,
we assume  that $M_{\rm SUSY}$ is relatively  heavy of order $\sim  2$ TeV.
In  calculating the Higgs couplings, we  have used  the following benchmark
values for  the soft SUSY-breaking parameters
\begin{equation}
  \label{Bench}
\begin{array}{l}
{\widetilde M}^2_Q = {\widetilde M}^2_L =  
{\widetilde M}^2_D = {\widetilde M}^2_E = 
{\widetilde M}^2_U\ = (1.7~{\rm TeV})^2,\\
A_u = A_d = A_e = 2.0~{\rm TeV}, \\
M_1 = M_2 = M_3 = 2.0~{\rm TeV}\; . 
\end{array}
\end{equation}
In addition, we take  $\mu=140~{\rm GeV}$ and $t_\beta=50$ throughout.
The Higgs-mediated  contribution to $B\to  X_s\gamma$ proceeds through
the  charged Higgs  boson  and the  predictions  of the  MSSM are  not
altered by the presence of the singlet Higgs bosons.  We have made use
of             the             public            code             {\tt
{\color{Red}CP}{\color{Blue}super}{\color{OliveGreen}H}}~\cite{CPsuperH}
to  check that  all points  considered  here are  consistent with  the
$2\sigma$ experimental bounds  on $B\to X_s\gamma$; the Higgs-mediated
contribution was found to be subdominant in the region of interest.

\subsection{Electroweak-scale Higgs singlets in the {\color{Red}MNSSM}}
\label{mnPhenom}

The renormalisable {\color{Red}MNSSM} superpotential is given by
\begin{equation}
\label{MNSSMsuperpotential}
{\mathcal W}_{\rm {\color{Red}MNSSM}} = {\mathcal W}_{\rm Yuk}
+\lambda \hat{S} \hat{H_1} \hat{H_2} +t_F \hat S\ ,
\end{equation}
where ${\mathcal W}_{\rm Yuk}$  represents the Yukawa couplings of the
MSSM superpotential.   The tadpole parameter $t_F$  and its associated
soft   SUSY-breaking   term  $t_S$   are   radiatively  generated   by
supergravity quantum effects from Planck-suppressed non-renormalisable
operators  in  the  K\"ahler  potential and  superpotential.  Discrete
$R$-symmetries, such as  $Z^R_5$ and $Z^R_7$, that are  imposed on the
theory postpones  the appearance of these operators  beyond the 5-loop
level, such that they are naturally suppressed of the order of $M_{\rm
SUSY}$ within a perturbative framework of supergravity (SUGRA).  For a
detailed  proof   of  the  argument,   see  the  first   reference  of
\cite{MNSSM}.

The tree-level Higgs sector of the {\color{Red}MNSSM} may be described
by the six parameters
\begin{equation}
t_\beta,\ M_a,\ \mu,\ \lambda,\ m_{12}^2,\ \frac{\lambda t_S}{\mu}\ ,
\end{equation}
where  $M_a$  is  the   would-be  MSSM  pseudoscalar  Higgs  mass  and
$m_{12}^2=\lambda t_F$.  In our numerical results we  also include the
dominant radiative corrections due to both (s)top and (s)bottom loops,
which are necessary to raise the mass of the SM-like Higgs boson above
the LEP II constraint of $114$~GeV.

Our  interest here  is in  the  effects of  light gauge-singlet  Higgs
bosons, with masses at or below the electroweak scale. Such a scenario
may be realised  by taking $M_a$ to be large, so  that the heavy Higgs
doublet fields  effectively decouple, and by  choosing $m_{12}^2$ such
that   the  mixing  between   the  light   CP-even  Higgs   bosons  is
suppressed. In the numerical estimates presented here, we take
\begin{equation}
 \label{MNSSMbench}
 M_a=1.5\ {\rm TeV},\quad m_{12}^2=(1.0\ {\rm TeV})^2,\quad 
\mu=140\ {\rm GeV},\quad \tan\beta=50, \quad
\lambda=0.3\ .
\end{equation}
This corresponds to a Higgs pseudoscalar mixing angle of $\cos\theta_A
\sim 0.17$.   The scalar Higgs  singlet mixes almost  exclusively with
the $\phi_1$ doublet, the  non-singlet fraction is approximately equal
for  both the  lightest Higgs  states  $H_1$ and  $A_1$.  These  light
particles are not constrained by  the direct search limits from LEP II
and  the  Tevatron due  to  their  highly  gauge-singlet nature,  with
$g_{H_{1,2} ZZ},\:  g_{H_1A_1Z}<0.05$.  The mass scale  of the singlet
Higgs    bosons   are    approximately    given   by    $\sqrt{\lambda
t_S/\mu}$.  Notice  that, in  the  absence  of strong  singlet-doublet
mixing  effects,  the  singlet   Higgs  scalar  and  pseudoscalar  are
approximately    degenerate   due    to   the    tree-level   mass-sum
rule~\cite{MNSSM}:
\begin{equation}
  \label{sumrule}
m^2_{H_1}\: +\: m^2_{H_2}\: +\:  m^2_{H_3}\ =\ M^2_Z\: +\: 
m^2_{A_1}\: +\:  m^2_{A_2}\; .
\end{equation}
Including dominant radiative corrections we find the masses of the
remaining Higgs bosons in this scenario to be
\begin{equation}
m_{H_2}=147\ {\rm GeV},\quad
m_{H_3}=1.52\ {\rm TeV},\quad
m_{A_2}=1.52\ {\rm TeV},
\end{equation}
so that the $H_2$ Higgs boson, which has SM-like couplings to the EW
gauge bosons, is well above the direct search limit set by LEP II.

\subsubsection{Effects on $\Delta M_{B_q}$}

\begin{figure}[t!]
\begin{center}
\includegraphics[scale=1.35]{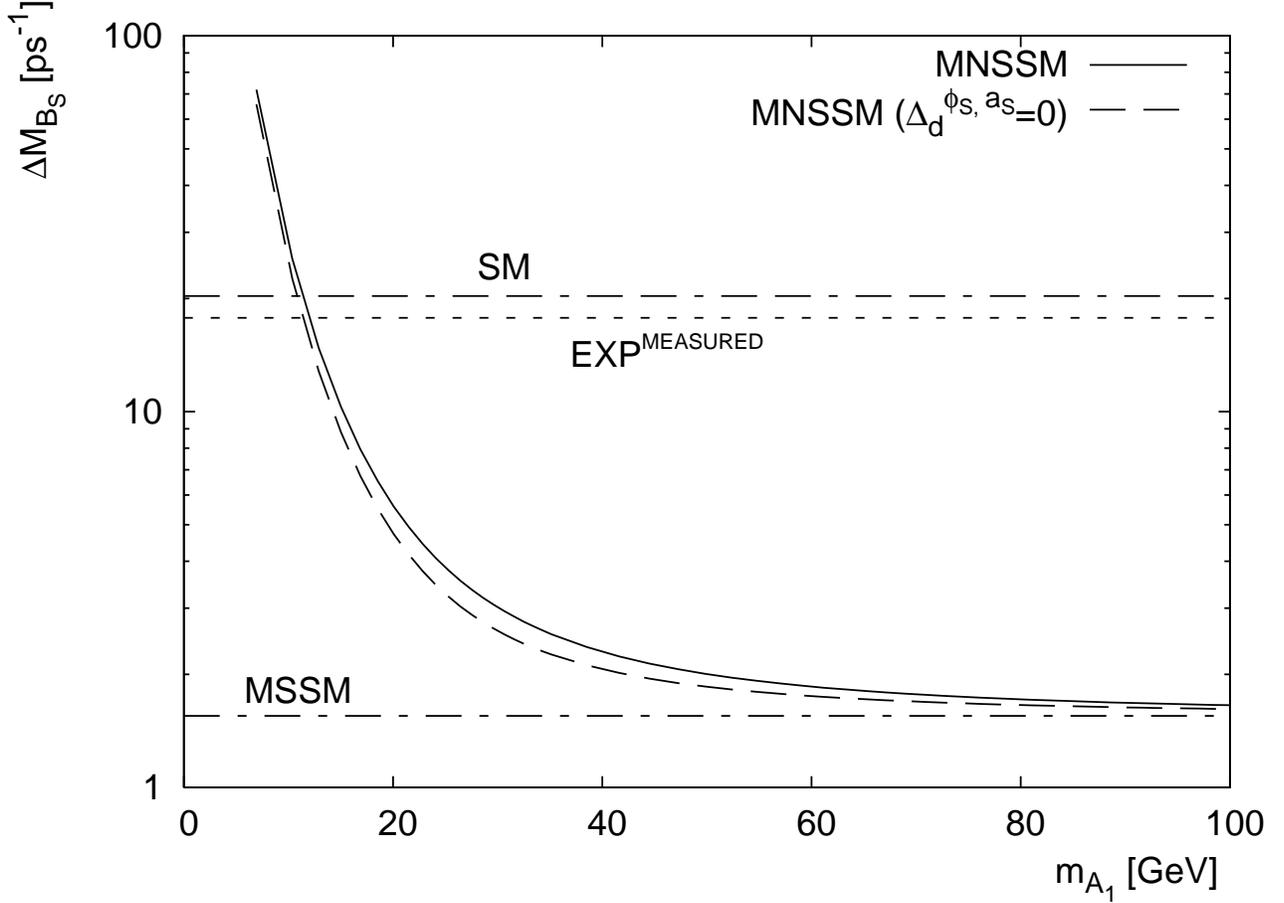}
\caption{\it The SUSY contribution to $\Delta M_{B_s}$ in units of
{\rm ps}$^{-1}$ as a function of the mass of the lightest pseudoscalar
$m_{A_1}$ in the {\color{Red}MNSSM}.  All parameters are taken as in~(\ref{Bench})
and~(\ref{MNSSMbench}).  The horizontal lines show the currently
measured value along with the SM and MSSM predictions
for corresponding values of $M_a$ and $\tan\beta$.}
\label{MNSSM_MBs}
\end{center}
\end{figure}

In  Fig.~\ref{MNSSM_MBs}  we show  the  SUSY  contribution to  $\Delta
M_{B_s}$  in  units  of  ps$^{-1}$,  as a  function  of  the  lightest
pseudoscalar  mass.   The upper  curve  fully  includes the  radiative
Yukawa couplings  of the Higgs  singlet fields whilst the  lower curve
neglects these corrections.  We observe that the SUSY contribution can
be larger  than the currently  observed value at large  $\tan\beta$ if
the singlet Higgs bosons are light.  For moderate values of $m_{A_1}$,
the effects of the  direct singlet Yukawa coupling remain significant,
enhancing the value of $\Delta M_{B_s}$ by around $25\%$.

\begin{figure}[t!]
\begin{center}
\includegraphics[scale=1.35]{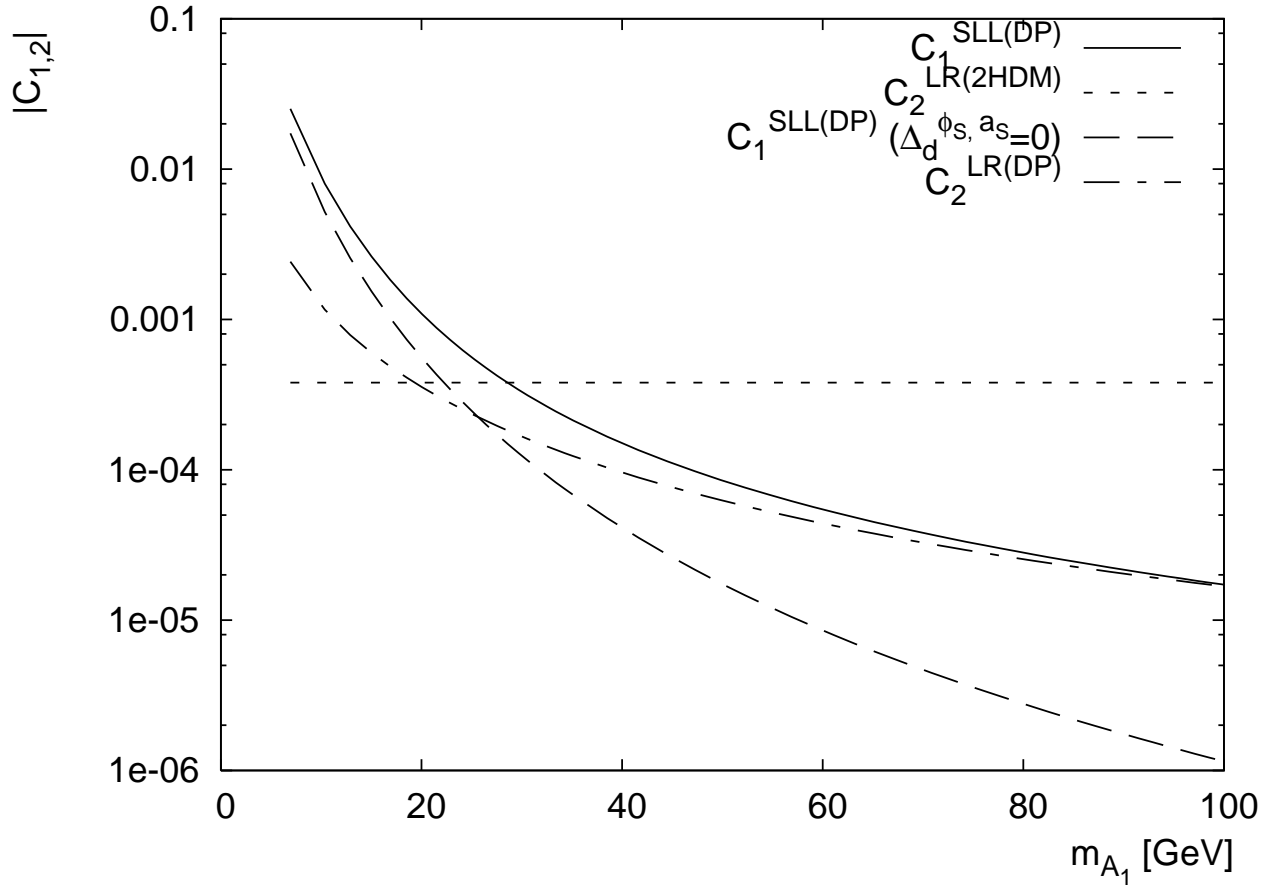}
\caption{\it The dominant Wilson coefficients contributing to $\Delta M_{B_s}$
as a function of the mass of the lightest pseudoscalar $m_{A_1}$ in the {\color{Red}MNSSM}.
The dashed curve shows the Wilson coefficient $C_1^{\rm SLL(DP)}$ neglecting the
threshold corrections for the gauge singlet Higgs boson.}
\label{WCoeffs}
\end{center}
\end{figure}

The dominant contribution to  $\Delta M_{B_s}$ for light singlet Higgs
bosons  is   due  to  the  Wilson   coefficient  $C_1^{\rm  SLL(DP)}$.
Considering  the  SHI approximation  for  the reduced  couplings~${\bf
g}^{L}_{H_1(A_1)\bar  b q}$, we  see that  the off-diagonal  terms are
given by
\begin{eqnarray}
{\bf g}^{L}_{H_1\bar b q} & = &
\left(t_\beta {\mathcal O}^H_{11}-\frac{v}{v_S}{\mathcal O}^H_{31}\right)
{\bf V}^\dag{\bf R}^{-1}{\bf V}\ ,\\
{\bf g}^{L}_{A_1\bar b q} & = &
i \left(t_\beta {\mathcal O}^A_{11}+\frac{v}{v_S}{\mathcal O}^A_{21}\right)
{\bf V}^\dag{\bf R}^{-1}{\bf V}\ ,
\end{eqnarray}
where  we  have neglected  the  small  contributions  of the  $\Phi_2$
component  next  to those  of  $\Phi_1$.   We  observe that  the  FCNC
couplings  of  the  lightest  scalar  and pseudoscalar  are  of  equal
magnitude,  if  threshold  corrections  to  the  singlet-Higgs  Yukawa
couplings  are  neglected and  ${\mathcal  O}^H_{11} \simeq  {\mathcal
O}^A_{11}$. In this  case, there is a cancellation  between the $H_1$-
and  the  $A_1$-mediated  contributions  to $C_1^{\rm  SLL(DP)}$.   An
analogous cancellation between  the heavy Higgs bosons $H$  and $A$ is
known to take  place in the MSSM  as well, as a consequence  of the PQ
symmetry which forbids  at the tree-level the dominant  $\Delta B = 2$
operators,  $(\bar{b}_R d_L)^2$  and  $(\bar{b}_R s_L)^2$,  associated
with $C_1^{\rm SLL(DP)}$ (see, e.g.~\cite{QCDfactorone}).

\begin{figure}[t!]
\begin{center}
\includegraphics[scale=1.35]{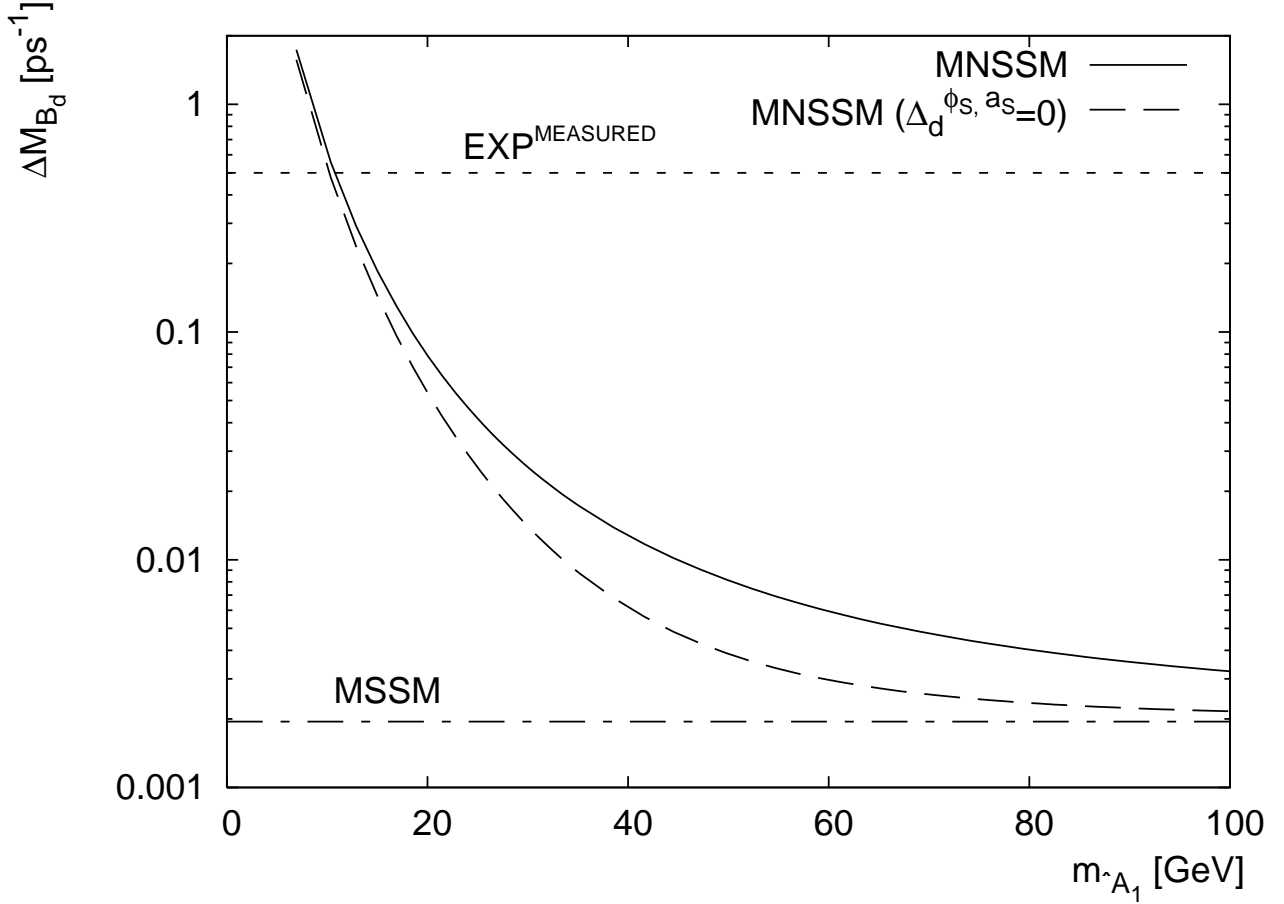}
\caption{\it The SUSY contribution to $\Delta M_{B_d}$ in units of
{\rm ps}$^{-1}$ as a function of the mass of the lightest pseudoscalar
$m_{A_1}$ in the {\color{Red}MNSSM}. The line conventions are as for
Fig.~\ref{MNSSM_MBs}.  All parameters are taken as in~(\ref{Bench})
and~(\ref{MNSSMbench}).  The upper horizontal line shows the currently
measured value and the lower horizontal line shows the MSSM prediction
for corresponding values of $M_a$ and $\tan\beta$.  We do not show the
SM prediction as the central value is close to the experimentally
observed splitting.}
\label{MNSSM_MBd}
\end{center}
\end{figure}

For very light singlets this  cancellation is dominantly broken by the
mass  splitting  between  $H_1$  and  $A_1$, typically  of  the  order
$m_{H_1}\sim  m_{A_1}+2$~GeV.  For  larger masses,  this  splitting is
negligible  and the  dominant  breaking is  instead  due to  threshold
effects  on the  singlet-Higgs Yukawa  couplings, which  are  found to
contribute to  the reduced  couplings with opposite  sign, due  to the
mixing  matrices.   The   dominant  Wilson  coefficients  for  $\Delta
M_{B_s}$ are  plotted in Fig.~\ref{WCoeffs}. We do  not plot $C_1^{\rm
SRR(DP)}$, which exhibits a  behaviour similar to $C_1^{\rm SLL(DP)}$,
but which is suppressed by a factor of $(m_s/m_b)^2$.

In Fig.~\ref{MNSSM_MBd} we show the corresponding plot for the contribution
to $\Delta M_{B_d}$. Again we observe that the SUSY contribution can exceed
the currently observed value, with the limit on $m_{A_1}$ at the same level
as that due to $\Delta M_{B_s}$. The effects of the singlet Yukawa coupling
are more pronounced at larger values of $m_{A_1}$ than in the case of
$\Delta M_{B_s}$, as the enhanced Wilson coefficient $C_1^{\rm SLL(DP)}$
remains dominant over the two-Higgs-doublet model contribution
$C_2^{\rm LR(2HDM)}$ by an order of magnitude here even for
$m_{A_1}\sim 100$~GeV.

\subsubsection{Effects on $\bar B^0_{d,s} \to \mu^+\mu^-$}

\begin{figure}[t!]
\begin{center}
\includegraphics[scale=1.35]{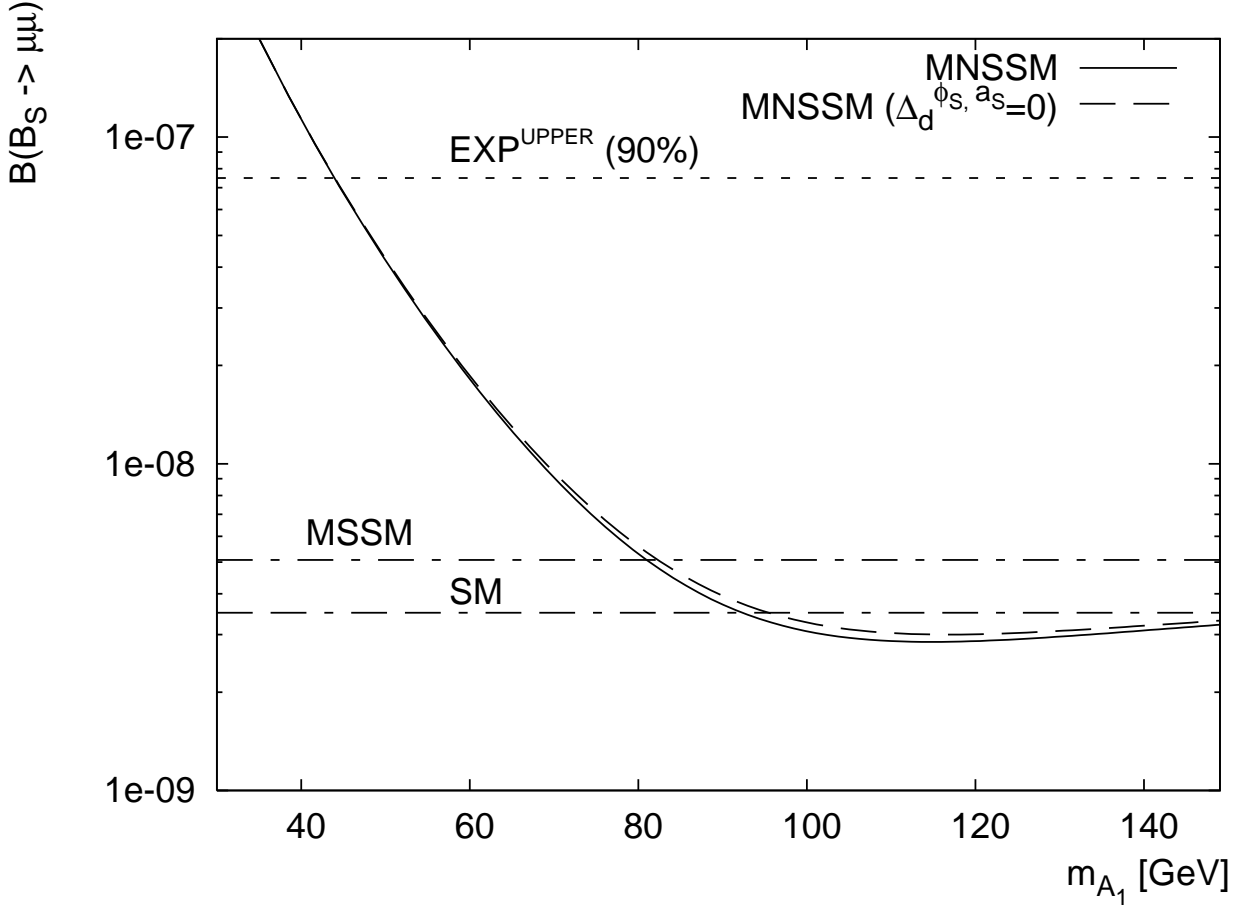}
\caption{\it The branching ratio ${\mathcal B}(B_s\to\mu^+\mu^-)$ as a
function of  the mass  of the lightest  pseudoscalar $m_{A_1}$  in the
{\color{Red}MNSSM}.  All  parameters  are  taken  as  in~(\ref{Bench})
and~(\ref{MNSSMbench}). The horizontal lines show the SM~\cite{BN} and
MSSM predictions  and the current  experimental upper limit  at $90\%$
C.L.}
\label{MNSSM_Bstomu}
\end{center}
\end{figure}

Figure~\ref{MNSSM_Bstomu} shows the branching ratio ${\mathcal
B}(B_s\to\mu^+\mu^-)$, as a function of $m_{A_1}$. As expected, the
prediction varies strongly with the mass of the light gauge singlet
Higgs bosons due to the $M_{H_1(A_1)}^{-4}$ dependence of the form
factors $F_{S(P)}^s$. The prediction exceeds the current bounds for
Higgs singlet masses below around $50$~GeV.  Since the branching ratio
in~(\ref{BranchingRateEquation}) depends only on the absolute values
of the form factors, there is no cancellation between the scalar and
pseudoscalar contributions.  At low values of $m_{A_1}$, including the
threshold corrections leads to a slight increase in $C_P$, which is
compensated for by a slight decrease in $C_S$ and the overall effect
is negligible.  At higher masses, $m_{A_1}\gsim 60$~GeV, there is
cancellation between the contributions due to $C_P$ and $C_{10}$.  In
this region the singlet threshold corrections produce a noticeable
shift.

\subsection{Light singlet Higgs pseudoscalar in the {\color{Blue}NMSSM}}
\label{NMPhenom}

The {\color{Blue}NMSSM} superpotential is given by
\begin{equation}
\label{NMSSMsuperpotential}
{\mathcal W}_{\rm {\color{Blue}NMSSM}} = {\mathcal W}_{\rm Yuk}
+\lambda \hat{S}\hat{H_1}\hat{H_2}+\frac{\kappa}{3}\hat{S}^3\ . 
\end{equation}
The corresponding soft SUSY-breaking terms are given by
\begin{equation}
{\mathcal L}^{\rm soft}_{\rm {\color{Blue}NMSSM}} =
\lambda A_\lambda S \Phi_1 \Phi_2
+\kappa A_\kappa S^3\ .
\end{equation}
In the limit  $A_{\lambda,\kappa} \to 0$, the scalar  potential of the
{\color{Blue}NMSSM}  possesses  an  additional  $U(1)$  symmetry  beyond
$U(1)_{\rm PQ}$.  This symmetry  results from an $R$-symmetry, denoted
by $U(1)_{\rm R}$, which is exact in the absence of soft SUSY-breaking
trilinear couplings.   Recall that the superpotential of  the NMSSM is
charged under U(1)$_{\rm  R}$ and so it differs  from $U(1)_{\rm PQ}$.
This $U(1)_{\rm R}$  is spontaneously broken when $S$  acquires a VEV,
so  that for small  values of  $A_{\lambda,\kappa}$ the  singlet Higgs
pseudoscalar  is  again  a  pseudo-Goldstone boson  and  is  naturally
expected    to    be   much    lighter    than    the   other    Higgs
fields~\cite{Rsymmetry,MNSSM}.

In order to  examine FCNC observables within such  a scenario, we take
the CP-odd mixing angle $\theta_A$  and the mass of the lightest Higgs
pseudoscalar   $m_{A_1}$   to  be   free   parameters   in  place   of
$A_{\lambda,\kappa}$, along with the couplings $\lambda$ and $\kappa$.
We use the following benchmark values throughout this section;
\begin{equation}
 \label{NMSSMbench}
 \lambda=0.4\,,\quad \kappa=-0.5\,, \quad \cos\theta_A=0.018\,, \quad
\mu=140\ {\rm GeV}\,,\quad \tan\beta=50\;,
\end{equation}
which are found to lead to the following masses for the remaining Higgs
bosons,
\begin{equation}
m_{H_1}=135\ {\rm GeV},\quad
m_{H_2}=357\ {\rm GeV},\quad
m_{H_3}=1.14\ {\rm TeV},\quad
m_{A_2}=1.14\ {\rm TeV}.
\end{equation}
The lightest CP-even Higgs in this scenario has SM-like couplings to
the EW gauge bosons.

For general values of $\kappa$, the CP-odd and CP-even singlets of the
{\color{Blue}NMSSM} are not constrained to have degenerate masses.  In
the scenario considered here, there  is only one light Higgs particle,
the  pseudo-Goldstone   boson  $A_1$.   Due  to  this,   there  is  no
cancellation  between the  dominant Higgs  field contributions  to the
Wilson  coefficients as  described in  Section~\ref{mnPhenom}  for the
{\color{Red}MNSSM}.  This forces us  to take a smaller singlet-doublet
mixing  parameter $\cos\theta_A$ in  order to  find phenomenologically
acceptable  results at  large values  of $\tan\beta$.   Note  that the
value of  $\cos\theta_A$ used here is  only $\sim 10\%$  of that found
for the {\color{Red}MNSSM} scenario considered previously and that the
contributions to ${\bf g}^{L(R)}_{A_1  \bar b q}$ from singlet-doublet
mixing and  direct threshold  corrections are therefore  comparable in
magnitude.

\subsubsection{Effects on $\Delta M_{B_q}$}

\begin{figure}[t!]
\begin{center}
\includegraphics[scale=1.35]{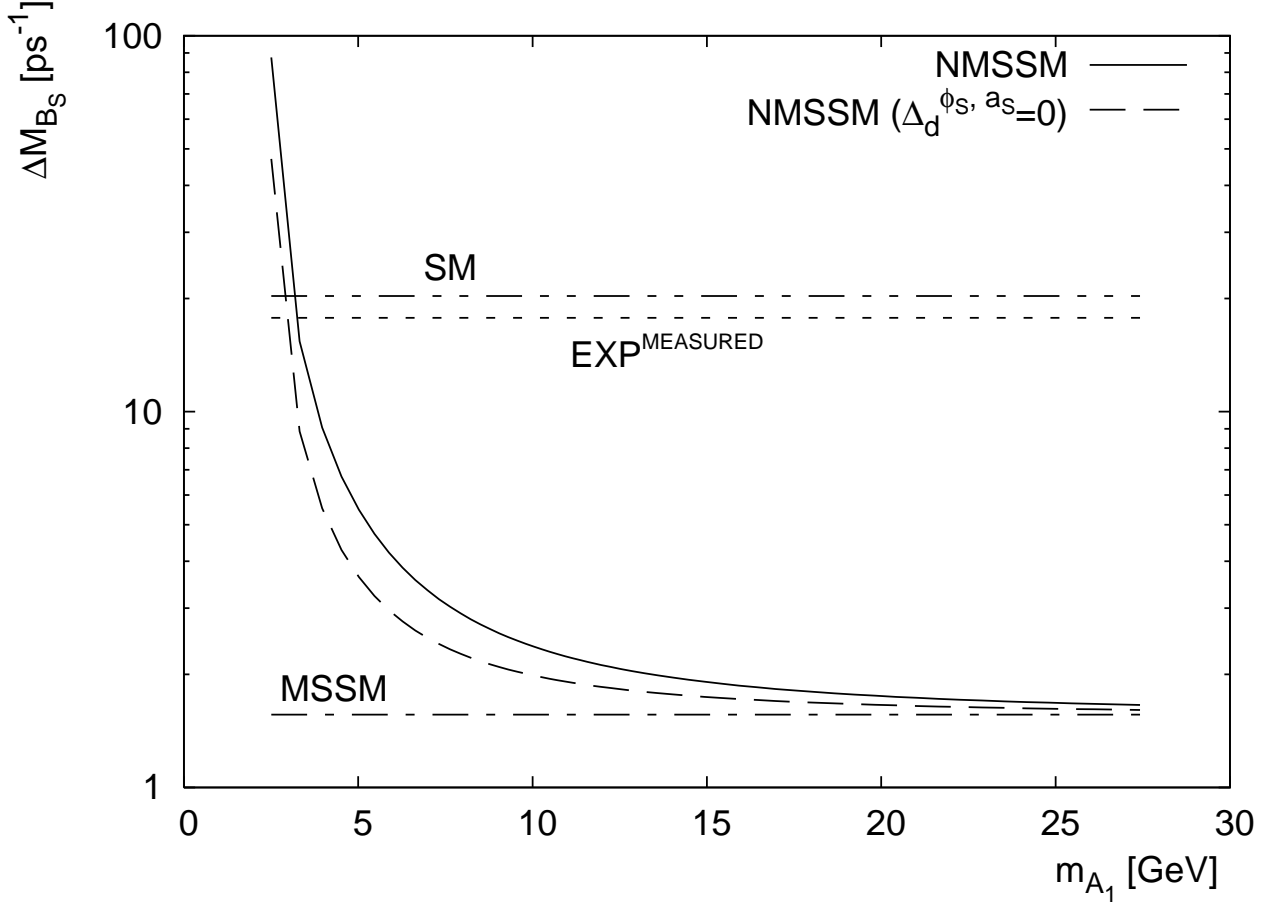}
\caption{\it  The SUSY contribution  to $\Delta  M_{B_s}$ in  units of
{\rm ps}$^{-1}$ as a function of the mass of the lightest pseudoscalar
$m_{A_1}$ in the {\color{Blue}NMSSM}.  The line conventions are as for
Fig.~\ref{MNSSM_MBs}.   All parameters  are taken  as in~(\ref{Bench})
and~(\ref{NMSSMbench}).   The horizontal lines show the currently
measured value along with the SM and MSSM predictions
for corresponding values of $M_a$ and $\tan\beta$.}
\label{NMSSM_MBs}
\end{center}
\end{figure}

In Fig.~\ref{NMSSM_MBs} we show the SUSY contribution to
$\Delta M_{B_s}$ as a function of the lightest Higgs
pseudoscalar mass.  In this scenario the SUSY contribution
also exceeds the currently measured value of $\Delta M_{B_s}$ for the
lightest values of $m_{A_1}$ at large values of $\tan\beta$.

We remark that  the contribution to $\Delta M_{B_s}$  in this scenario
of   the   {\color{Blue}NMSSM}  is   rather   smaller   than  in   the
{\color{Red}MNSSM} scenario  discussed in Section~\ref{mnPhenom}.  The
leading contribution  in both cases  is due to the  Wilson coefficient
$C_1^{\rm SLL(DP)}$.   As discussed above,  in the {\color{Blue}NMSSM}
there  is no cancellation  between dominant  CP-odd and  CP-even Higgs
fields and so  we have taken the FCNC Yukawa couplings  to be an order
of magnitude  smaller than those  considered in the  previous section.
As  a result  of this,  the contribution  of the  singlet pseudoscalar
become negligible  here for $m_{A_1}\gsim  25$~GeV and we  recover the
MSSM prediction.

\begin{figure}[t!]
\begin{center}
\includegraphics[scale=1.35]{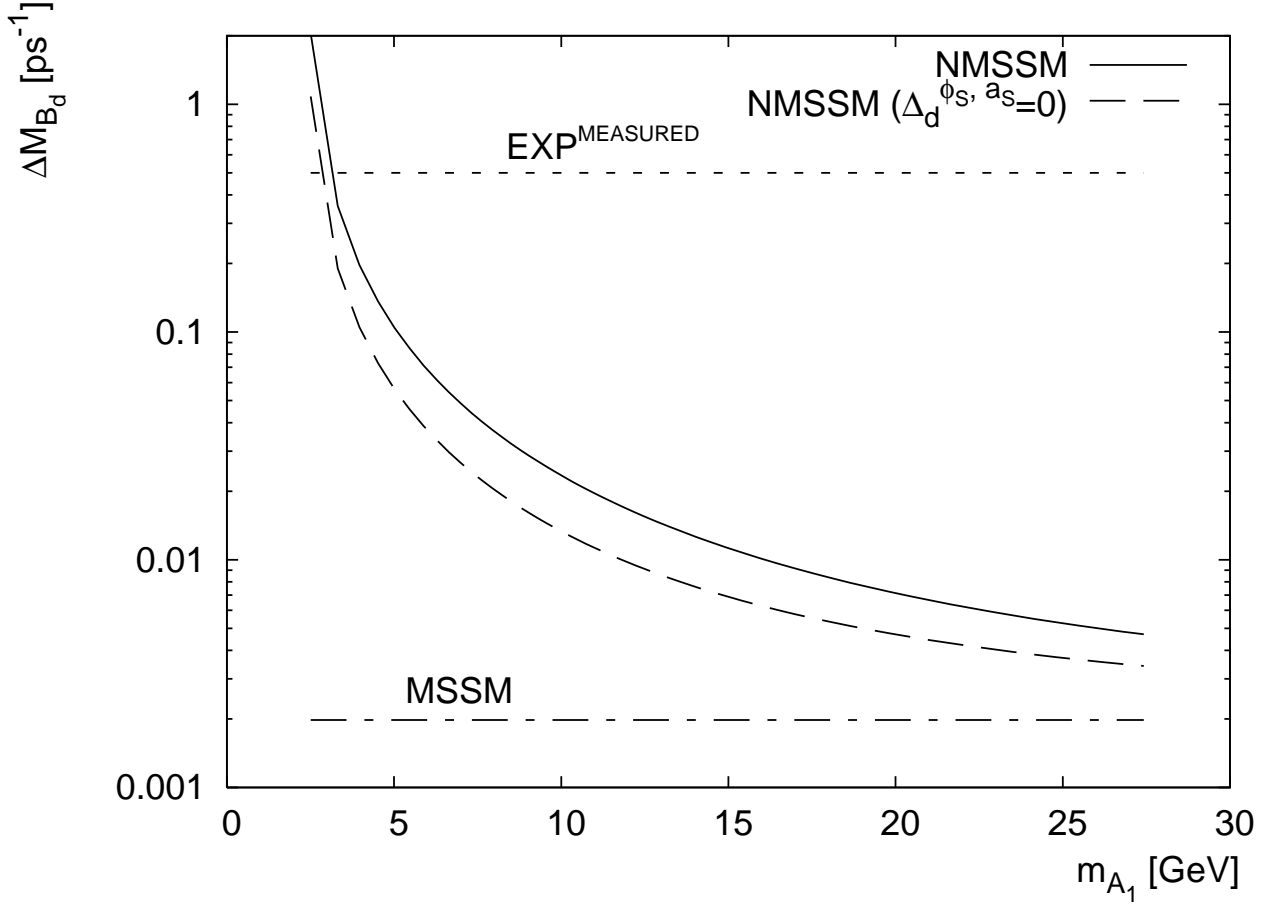}
\caption{\it The SUSY contribution to $\Delta M_{B_d}$ in units of
{\rm ps}$^{-1}$ as a function of the mass of the lightest pseudoscalar
$m_{A_1}$ in the {\color{Blue}NMSSM}. The line conventions are as for
Fig.~\ref{MNSSM_MBs}.  All parameters are taken as in~(\ref{Bench})
and~(\ref{NMSSMbench}).  The upper horizontal line shows the currently
measured value and the lower horizontal line shows the MSSM prediction
for corresponding values of $M_a$ and $\tan\beta$.  We do not show the
SM prediction as the central value is close to the experimentally
observed splitting.}
\label{NMSSM_MBd}
\end{center}
\end{figure}

Figure~\ref{NMSSM_MBd} shows the corresponding predictions for $\Delta
M_{B_d}$.   As  in   the  {\color{Red}MNSSM}  scenario  considered  in
Section~\ref{mnPhenom} the bounds from $B_d  - \bar B_d$ mixing are at
the same level  as those from the $B_s$  mesons, although the relative
importance of  the singlet threshold  corrections at larger  masses is
greater due  to the continuing  dominance of $C_1^{\rm  SLL(DP)}$ over
$C_2^{\rm LR(DP)}$.   In particular, the contributions  of the singlet
pseudoscalar are not negligible at  the same low masses as for $\Delta
M_{B_s}$.

\subsubsection{Effects on $\bar B^0_{s} \to \mu^+\mu^-$}

\begin{figure}[t!]
\begin{center}
\includegraphics[scale=1.35]{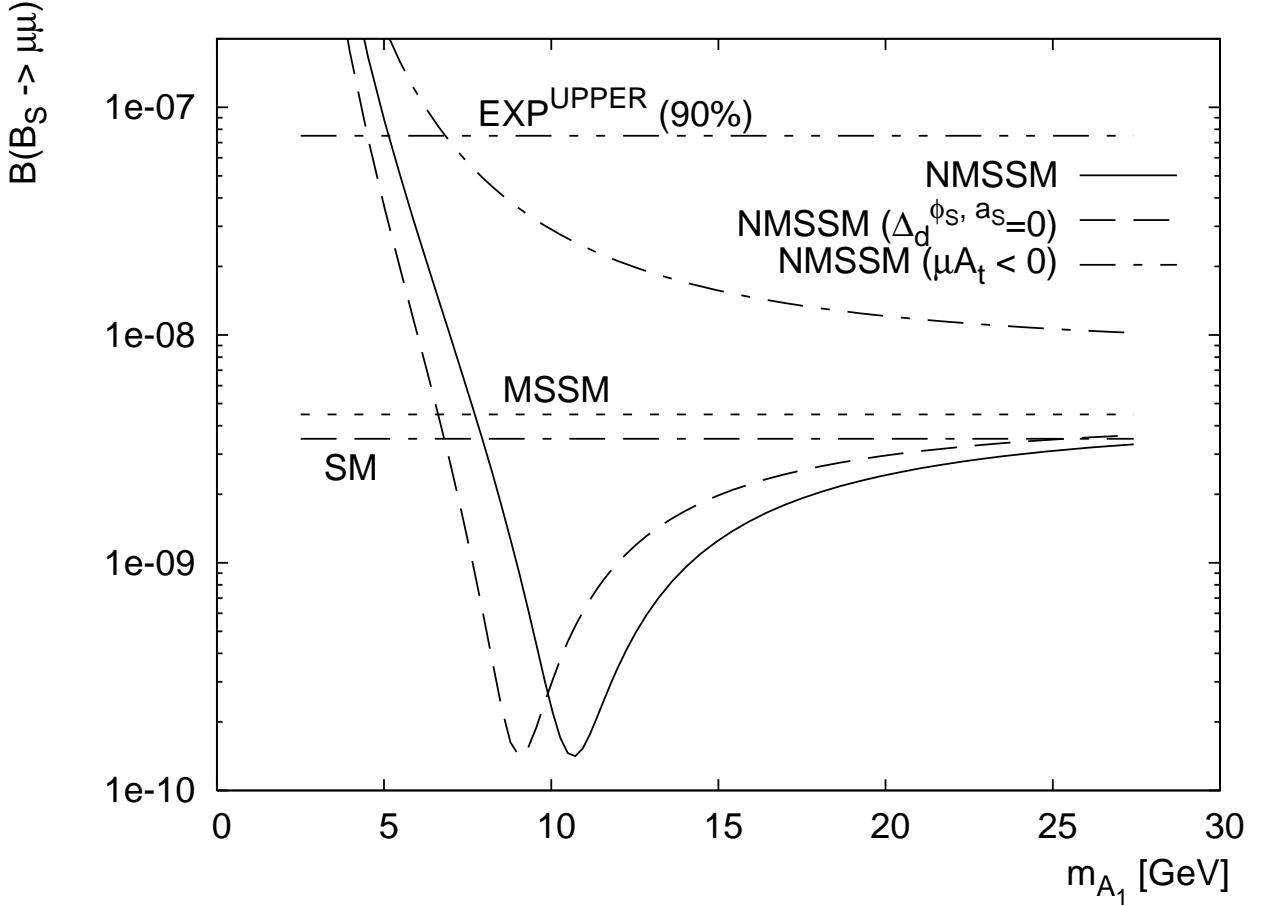}
\caption{\it The branching ratio ${\mathcal B}(B_s\to\mu^+\mu^-)$ as a
function of  the mass of  the lightest pseudoscalar, $m_{A_1}$  in the
{\color{Blue}NMSSM}. The solid line includes threshold corrections for
the gauge singlet Higgs bosons,  the dashed curve neglects these.  All
parameters are taken  as in~(\ref{Bench}) and~(\ref{NMSSMbench}).  The
dot-dashed curve  shows the NMSSM prediction,  including all threshold
corrections,  but with  $A_u=-2$~TeV.  The  horizontal lines  show the
SM~\cite{BN}  and the  MSSM predictions  and the  current experimental
upper limit at $90\%$ C.L. respectively.}
\label{NMSSM_Bstomu}
\end{center}
\end{figure}

Figure~\ref{NMSSM_Bstomu} shows the branching ratio ${\mathcal B}
(B_s\to \mu^+\mu^-)$ as a function of the lightest pseudoscalar mass
in the {\color{Blue}NMSSM}. For small values of $m_{A_1}$ the
prediction exceeds the current $90\%$ confidence limits at large
$\tan\beta$.  For a heavier singlet Higgs pseudoscalar, we observe
that inclusion of the singlet threshold corrections leads to a
significant reduction in the branching ratio, which becomes suppressed
by more than an order of magnitude compared to the SM prediction.  In
this mass range the contribution of the light pseudoscalar interferes
destructively with the SM-like diagrams.  It should be noted that such
a large suppression is not possible in either the MSSM or
{\color{Red}MNSSM}, since in both these models tree-level mass-sum
rules prevent the appearance of isolated CP-odd Higgs bosons.

The  leading effects  of the  neutral  Higgs bosons  on ${\mathcal  B}
(B_s\to \mu^+\mu^-)$  in both the  MSSM and {\color{Red}MNSSM}  may be
understood  by including only  the contributions  due to  the dominant
scalar/pseudoscalar pair (i.e.  $H,A$ in  the MSSM or $H_1,A_1$ in the
above  considered limit  of  the {\color{Red}MNSSM})  and by  assuming
equal masses and  effective Yukawa couplings (this is  accurate to the
order of a few percent).  In  this approximation, it can be shown that
the  absolute minimum  of  ${\mathcal B}(B_s\to  \mu^+\mu^-)$ in  both
these models is $1/2 \times {\mathcal B}(B_s\to \mu^+\mu^-)_{\rm SM}$.

When such a cancellation  is effective in the {\color{Blue}NMSSM}, the
absolute minimum  of ${\mathcal B}(B_s\to  \mu^+\mu^-)$ is due  to the
contribution of the heavy CP-even Higgs doublet, $H_3\sim\phi_1$.  The
exact value of ${\mathcal B}(B_s\to \mu^+\mu^-)$ at the minimum scales
as $M_{H_3}^{-4}$  and the branching ratio will  be further suppressed
as $H_3$ and $A_2$ become heavier.  As can be seen from the dot-dashed
curve in  Fig.~\ref{NMSSM_Bstomu}, the suppression is  found to vanish
when $\mu A_u$ flips sign,  i.e.~for $A_u = -2$~TeV, since the leading
SM and SUSY contributions interfere constructively in this case.

\section{Conclusions}

We  have  shown  that  singlet  Higgs bosons  generically  present  in
extensions  of   the  MSSM   can  contribute  significantly   to  FCNC
observables at large $\tan\beta$, both through their tree-level mixing
with the Higgs doublet fields and through 1-loop quantum effects.  For
very light  Higgs singlets, the  contributions may exceed even  by one
order of magnitude the  current experimental limits on FCNC processes.
This allows us to place severe constraints on the mass of the lightest
scalar state in the theory at large values of $\tan\beta$.

In the  {\color{Red}MNSSM}, the scalar and  pseudoscalar singlet Higgs
fields are constrained to  be approximately degenerate by a tree-level
mass sum  rule~\cite{MNSSM}.  Their contributions  to $B$-meson mixing
through  the  dominant Wilson  coefficient  $C_1^{\rm SLL}$  interfere
destructively,  analogously to  the  contributions of  the MSSM  heavy
Higgs bosons.  At moderate to large values of the common Higgs singlet
mass,  the direct  coupling plays  an important  role in  lifting this
cancellation,  leading  to a  large  enhancement  of $C_1^{\rm  SLL}$.
Although   cancellations   between   the  pseudoscalar   and   SM-like
contributions to the  branching ratio $B_s\to\mu^+\mu^-$ are possible,
the quasi-degenerate scalar Higgs singlet produces a screening effect.
Suppression of the branching ratio  below the SM prediction is limited
to at most a factor of $\sim 1/2$ at leading order.

The situation  in the  {\color{Blue}NMSSM} is different.   The CP-even
singlet  remains   heavy  close  to  the   $R$-symmetric  limit.   The
contribution of  the singlet pseudoscalar can  therefore easily exceed
the  experimental limits at  large values  of $\tan\beta$,  unless the
double-singlet mixing is constrained to  be small.  Due to the absence
of an accompanying CP-even  Higgs boson, the cancellation between SUSY
and SM-like  contribution to the  branching ratio $B(B_s\to\mu^+\mu^-)$
can be highly  efficient in this model, where  suppression can be more
than one  order of magnitude relative  to the SM  prediction.  Such an
effect is not possible within the MSSM at large $\tan\beta$.  Possible
discovery of  SUSY partners at  the LHC along with  non-observation of
the  decay   $B_s\to\mu^+\mu^-$  would  strongly   point  towards  the
{\color{Blue}NMSSM}  and  searches  for light  pseudoscalar  particles
should become a high priority,  should such a situation present itself
in the near future.

The  present work  was carried  out  within the  framework of  minimal
flavour violation.  However, within a general renormalisable framework
of the models we have been  studying here, new sources of CP violation
could  be  present amongst  the  soft  SUSY-breaking  terms.  This  is
expected to  lead to new  CP-violating threshold corrections  for both
the  singlet  and  doublet   Higgs  fields.   Additionally,  both  the
{\color{Blue}NMSSM}  and the  {\color{Red}MNSSM}  exhibit explicit  CP
violation in  the tree-level Higgs potential.  These  effects may lead
to significant enhancements of the signals considered here, along with
contributions to  other observables, such as  electric dipole moments.
We plan to report progress on these issues in the near future.

\section*{Acknowledgments}

This  work  was  partially  supported  by  the  STFC  research  grant:
PP/D000157/1.

\newpage
\begin{appendix}
\setcounter{equation}{0}
\def\theequation{\Alph{section}.\arabic{equation}}

\section{The Chargino-Neutralino Propagator Matrix}

In~(\ref{matrixintegral}), we need to evaluate the different left- and
right-handed chiral  components of the  chargino-neutralino propagator
matrix:
\begin{equation}
  \label{PropMC}
\Delta_C (\not\! k) \ =\ {1\over \slash k {\bf 1}_9-{\bf M}_C P_L
-{\bf M}_C^\dag P_R}\ .
\end{equation}
To do  so, it proves more  convenient to use  the chiral representation
for~(\ref{PropMC}) and write $\Delta_C (\not\! k)$ as
\begin{equation}
  \label{PropMCchiral}
\Delta_C (\not\! k) \ =\
\left(\begin{array}{cc}
-{\bf M}_C {\bf 1}_2 & {\bf 1}_9 {\bf k}  \\
{\bf 1}_9 {\bf\bar k} & -{\bf M}_C^\dag {\bf 1}_2\end{array}\right)^{-1}\ .
\end{equation}
In the above, we have defined ${\bf k}=k_\mu \sigma^\mu$ and $\bar{\bf
k}=k_\mu   \bar{\sigma}^\mu$,   where   $\sigma^\mu=   ({\bf   1}_2,\,
\mbox{\boldmath          $\sigma$})$,          $\bar{\sigma}^\mu=({\bf
1}_2,\,-\mbox{\boldmath  $\sigma$})$ and  {\boldmath $\sigma$}  are the
usual Pauli matrices.  Note  that ${\bf k\bar k}={\bf\bar k k}=k^2{\bf
1_2}$.

Our  calculational task  is equivalent  to finding  the inverse  of an
invertible, block matrix~$M$,
\begin{equation}
M=\left(\begin{array}{cc}
A & B \\
C & D \end{array}\right)\ ,
\end{equation}
where $A,\, B,\,  C$ and $D$ are $m\times  m$-, $m\times n$-, $n\times
m$- and  $n\times n$-dimensional matrices, respectively.  To this end,
we denote the inverse of $M$ by
\begin{equation}
M^{-1}
=
\left(\begin{array}{cc}
\alpha & \beta \\
\gamma & \delta \end{array}\right)\ .
\end{equation}
Correspondingly, $\alpha, \ \beta,\  \gamma$ and $\delta$ are matrices
of dimensionality  $m\times m$, $m\times n$, $n\times  m$ and $n\times
n$.  We may  now use the condition $MM^{-1}={\bf  1}_{(n+m)}$ to write
the four constraints,
\begin{eqnarray}
\label{appen_one}
A\alpha + B\gamma & = & {\bf 1}_m\ ,\\
\label{appen_two}
C\alpha + D\gamma & = & {\bf 0}_{n\times m}\ ,\\ 
\label{appen_three}
A\beta + B\delta & = & {\bf 0}_{m\times n}\ ,\\
\label{appen_four}
C\beta + D\delta & = & {\bf 1}_n\ .
\end{eqnarray}
Conditions (\ref{appen_two}) and (\ref{appen_three}) imply that
\begin{eqnarray}
\gamma & = & -D^{-1}C\alpha\ ,\\
\beta & = & -A^{-1}B\delta\ .
\end{eqnarray}
Combining these with~(\ref{appen_one}) and~(\ref{appen_four}), we may derive
\begin{eqnarray}
\alpha & = & \left(A-BD^{-1}C\right)^{-1}\ ,\\
\delta & = & \left(D-CA^{-1}B\right)^{-1}\ ,
\end{eqnarray}
so that the inverse matrix $M^{-1}$ may be written as
\begin{equation}
\label{appen_inverse}
M^{-1}\
 =\
\left(\begin{array}{cc}
\left(A-BD^{-1}C\right)^{-1} & -A^{-1}B\left(D-CA^{-1}B\right)^{-1}\\
-D^{-1}C\left(A-BD^{-1}C\right)^{-1} & \left(D-CA^{-1}B\right)^{-1}
\end{array}\right)\ .
\end{equation}
One should  observe here that  the expression $M^{-1}$ is  not unique.
Specifically, we could  have used $M^{-1}M = {\bf  1}_{m+n}$ to obtain
equations   analogous  to   (\ref{appen_one})--(\ref{appen_four})  and
express $M^{-1}$ in a form which, although less compact, relies on the
inverses of  two combinations  of submatrices only,  e.g.~$D^{-1}$ and
$\left(A-BD^{-1}C\right)^{-1}$.  Such expressions are more efficiently
implemented  in  numerical routines.  However,  all these  alternative
forms are equivalent to each~other.

We  now  return  to  $\Delta_C (\not\!   k)$  in~(\ref{PropMCchiral}).
Using~(\ref{appen_inverse}), the chargino-neutralino propagator matrix
may now be written down as
\begin{equation}
  \label{DeltaC}
\Delta_C (\not\! k) \ =\
\left(\begin{array}{cc}
{\bf M}_C^\dag\left(k^2 {\bf 1}_9 - {\bf M}_C {\bf M}_C^\dag\right)^{-1}{\bf 1}_2
 &
\left(k^2{\bf 1}_9- {\bf M}^\dag_C {\bf M}_C\right)^{-1}{\bf k}\\
\left(k^2{\bf 1}_9 - {\bf M}_C{\bf M}_C^\dag\right)^{-1}{\bf\bar k}
&
{\bf M}_C \left(k^2{\bf 1}_9 - {\bf M}^\dag_C
{\bf M}_C\right)^{-1}{\bf 1}_2\end{array}\right)\ ,
\end{equation}
where we have used the  fact that the only non-zero commutator amongst
the  submatrices  is  $\left[A,D\right]$  which simplifies  the  above
expression. From~(\ref{DeltaC}), we may now obtain the different block
elements,  by   appropriately  acting  with   left-  and  right-handed
projection operators  $P_{L.R}$. For example, for  the expression that
occurs in~(\ref{matrixintegral}), we find that
\begin{equation}
  \label{PLDCPL}
P_L\, \Delta_C (\not\! k)\, P_L \ =\
{\bf M}_C^\dag \left(k^2{\bf 1}_9 - {\bf M}_C {\bf M}_C^\dag
\right)^{-1} {\bf 1}_2\ ,
\end{equation}
where  we  have  suppressed  the  vanishing elements  on  the  RHS  of
(\ref{PLDCPL}).

Using~(\ref{PLDCPL}),  we may  numerically integrate  expressions such
as~(\ref{matrixintegral})  without any need  to first  diagonalise the
mass matrices appearing  in the propagators.  At each  sample point in
the integral, the propagator matrices are numerically inverted and the
matrix  element  relevant  to  the  process  is  selected,  i.e.   the
propagator  matrix  element representing  the  transition between  the
required  electroweak   eigenstates.   In  this  way,   we  avoid  the
introduction  of  cumbersome  mixing  matrices  for  the  squarks  and
charginos       in       the       analytic       expressions       of
Section~\ref{EffectiveLagrangian}.

Furthermore,  by   working  with   all  propagator  matrices   in  the
electroweak  basis,  the  calculation  of  the  derivatives  appearing
in~(\ref{hlet})  is  greatly  simplified.   The  expression  for  each
derivate of the  propagators in the matrix form  contains no more than
two terms.  By contrast, if  we were to diagonalise the mass matrices,
we  would  find  expressions  containing  three terms  for  each  mass
eigenstate (coming  from applying the  Leibniz rule to the  product of
Higgs-dependant masses  with initial and final  mixing matrices), many
of  which typically require  further expansion  before one  could cast
them into a useful compact form.

\end{appendix}

\newpage

\end{document}

%% file: def_colors
\definecolor{GreenYellow}   {cmyk}{0.15,0,0.69,0}
\definecolor{Yellow}        {cmyk}{0,0,1,0}
\definecolor{Goldenrod}     {cmyk}{0,0.10,0.84,0}
\definecolor{Dandelion}     {cmyk}{0,0.29,0.84,0}
\definecolor{Apricot}       {cmyk}{0,0.32,0.52,0}
\definecolor{Peach}         {cmyk}{0,0.50,0.70,0}
\definecolor{Melon}         {cmyk}{0,0.46,0.50,0}
\definecolor{YellowOrange}  {cmyk}{0,0.42,1,0}
\definecolor{Orange}        {cmyk}{0,0.61,0.87,0}
\definecolor{BurntOrange}   {cmyk}{0,0.51,1,0}
\definecolor{Bittersweet}   {cmyk}{0,0.75,1,0.24}
\definecolor{RedOrange}     {cmyk}{0,0.77,0.87,0}
\definecolor{Mahogany}      {cmyk}{0,0.85,0.87,0.35}
\definecolor{Maroon}        {cmyk}{0,0.87,0.68,0.32}
\definecolor{BrickRed}      {cmyk}{0,0.89,0.94,0.28}
\definecolor{Red}           {cmyk}{0,1,1,0}
\definecolor{OrangeRed}     {cmyk}{0,1,0.50,0}
\definecolor{RubineRed}     {cmyk}{0,1,0.13,0}
\definecolor{WildStrawberry}{cmyk}{0,0.96,0.39,0}
\definecolor{Salmon}        {cmyk}{0,0.53,0.38,0}
\definecolor{CarnationPink} {cmyk}{0,0.63,0,0}
\definecolor{Magenta}       {cmyk}{0,1,0,0}
\definecolor{VioletRed}     {cmyk}{0,0.81,0,0}
\definecolor{Rhodamine}     {cmyk}{0,0.82,0,0}
\definecolor{Mulberry}      {cmyk}{0.34,0.90,0,0.02}
\definecolor{RedViolet}     {cmyk}{0.07,0.90,0,0.34}
\definecolor{Fuchsia}       {cmyk}{0.47,0.91,0,0.08}
\definecolor{Lavender}      {cmyk}{0,0.48,0,0}
\definecolor{Thistle}       {cmyk}{0.12,0.59,0,0}
\definecolor{Orchid}        {cmyk}{0.32,0.64,0,0}
\definecolor{DarkOrchid}    {cmyk}{0.40,0.80,0.20,0}
\definecolor{Purple}        {cmyk}{0.45,0.86,0,0}
\definecolor{Plum}          {cmyk}{0.50,1,0,0}
\definecolor{Violet}        {cmyk}{0.79,0.88,0,0}
\definecolor{RoyalPurple}   {cmyk}{0.75,0.90,0,0}
\definecolor{BlueViolet}    {cmyk}{0.86,0.91,0,0.04}
\definecolor{Periwinkle}    {cmyk}{0.57,0.55,0,0}
\definecolor{CadetBlue}     {cmyk}{0.62,0.57,0.23,0}
\definecolor{CornflowerBlue}{cmyk}{0.65,0.13,0,0}
\definecolor{MidnightBlue}  {cmyk}{0.98,0.13,0,0.43}
\definecolor{NavyBlue}      {cmyk}{0.94,0.54,0,0}
\definecolor{RoyalBlue}     {cmyk}{1,0.50,0,0}
\definecolor{Blue}          {cmyk}{1,1,0,0}
\definecolor{Cerulean}      {cmyk}{0.94,0.11,0,0}
\definecolor{Cyan}          {cmyk}{1,0,0,0}
\definecolor{ProcessBlue}   {cmyk}{0.96,0,0,0}
\definecolor{SkyBlue}       {cmyk}{0.62,0,0.12,0}
\definecolor{Turquoise}     {cmyk}{0.85,0,0.20,0}
\definecolor{TealBlue}      {cmyk}{0.86,0,0.34,0.02}
\definecolor{Aquamarine}    {cmyk}{0.82,0,0.30,0}
\definecolor{BlueGreen}     {cmyk}{0.85,0,0.33,0}
\definecolor{Emerald}       {cmyk}{1,0,0.50,0}
\definecolor{JungleGreen}   {cmyk}{0.99,0,0.52,0}
\definecolor{SeaGreen}      {cmyk}{0.69,0,0.50,0}
\definecolor{Green}         {cmyk}{1,0,1,0}
\definecolor{ForestGreen}   {cmyk}{0.91,0,0.88,0.12}
\definecolor{PineGreen}     {cmyk}{0.92,0,0.59,0.25}
\definecolor{LimeGreen}     {cmyk}{0.50,0,1,0}
\definecolor{YellowGreen}   {cmyk}{0.44,0,0.74,0}
\definecolor{SpringGreen}   {cmyk}{0.26,0,0.76,0}
\definecolor{OliveGreen}    {cmyk}{0.64,0,0.95,0.40}
\definecolor{RawSienna}     {cmyk}{0,0.72,1,0.45}
\definecolor{Sepia}         {cmyk}{0,0.83,1,0.70}
\definecolor{Brown}         {cmyk}{0,0.81,1,0.60}
\definecolor{Tan}           {cmyk}{0.14,0.42,0.56,0}
\definecolor{Gray}          {cmyk}{0,0,0,0.50}
\definecolor{Black}         {cmyk}{0,0,0,1}
\definecolor{White}         {cmyk}{0,0,0,0}